\documentclass[manuscript,nonacm]{acmart}

\AtBeginDocument{%
  }
\usepackage{marvosym}

\begin{document}

\title{What Do We Mean When We Say `F*ck AI'?}
\subtitle{Moral Panic, Semantic Indeterminacy, and the Reification of Techno-Capitalism}

\author{Sasha Luccioni}
\authornote{Both authors contributed equally to this research.}
\authornotemark[1]
\affiliation{%
  \institution{Sustainable AI Group}
  \city{Montr{\'e}al}
  \country{Canada}}
  \correspondingauthor
\email{}

\author{Travis LaCroix}
\authornotemark[2]
\affiliation{%
  \institution{Durham University}
  \city{Durham}
  \country{United Kingdom}
}

\renewcommand{\shortauthors}{Luccioni and LaCroix}

\begin{abstract}
    Academic and (increasingly) public resistance to artificial intelligence (AI) often blurs the line between legitimate critique of techno-capitalist exploitation and (inflated) existential risk. Moreover, discourse around AI can sometimes resemble a ``moral panic'' where an apparently clear and identifiable scapegoat is targeted in lieu of the complex systemic problems that are causing (genuine and justified) social anxieties.  However, AI is not a monolith. In this paper, we explore the environmental, political, and socio-technical consequences of the semantic indeterminacy of discourse about ``AI'' in conjunction with the increasing moral panic encompassing these technologies. We argue that since the umbrella term ``AI'' lacks a coherent, singular referent, the moral panic about AI inadvertently reifies the concept of an autonomous, singular agent with inherent and often autonomous tendencies, which aligns with corporate narratives. Hence, well-intentioned critique, activist pushback, and public protest paradoxically amplifies hype-based discussions of AI, thus validating the ``threat'' narrative that tech leaders deploy to exaggerate their products' capabilities and market value, further consolidating and centralising their political and socio-economic power. 
    Moreover, if ``AI'' does not refer to a unified class of technologies, then policy intended to govern ``AI'' will systematically (and necessarily) fail. 
    We conclude by proposing ways forward to improve AI literacy, enforce accountability, and repurpose existing regulation in order to imagine a more positive AI future.
\end{abstract}


    \begin{teaserfigure}
        \centering
        \includegraphics[width=0.55\textwidth]%
        {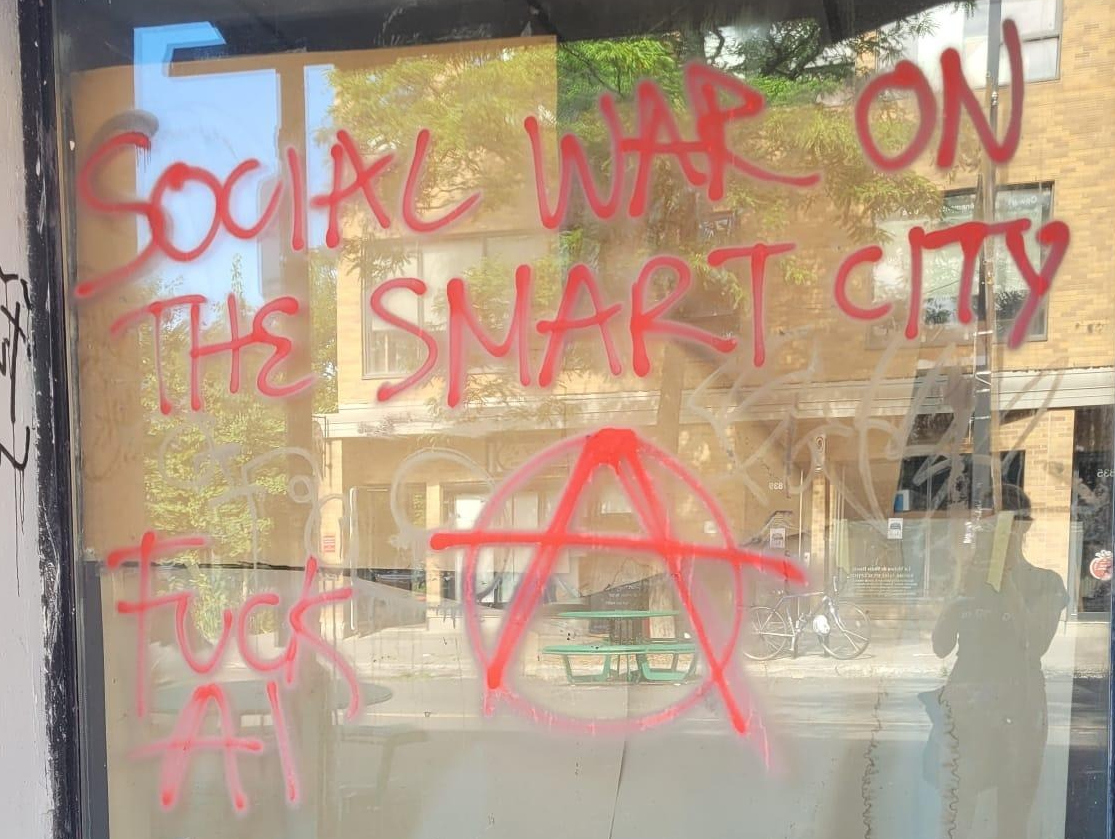}
        \caption{Graffiti reading ``SOCIAL WAR ON THE SMART CITY'' and ``FUCK AI'', alongside the circle-A symbol (\CircledA)\ 
        on the window of Montr{\'e}al building, taken by Sasha Luccioni. See Appendix~\ref{sec:A} for additional examples.}
    \Description{A photo showing red, French graffiti on a glass window, which reads SOCIAL WAR ON THE SMART CITY'' and ``FUCK AI'', alongside the circle-A symbol.}
    \label{fig:Graffiti1}
    \end{teaserfigure}

\maketitle

\section{Introduction} \label{sec:intro}

    Recent years have seen the popularisation of increasingly conflicting narratives about artificial intelligence (AI). On the one hand, the CEOs of major AI companies have promised that these technologies will usher in an  ``age of abundance''~\cite{gulfnews2026}, cure cancer~\cite{altman2025} (or, all diseases)~\cite{ET-Now-2025}, ``fix'' climate crises~\cite{altman2024}, and drastically accelerate economic growth~\cite{amodei2026}, creating an economy of leisure~\cite{Klein-2026}. In effect, ``more intelligence'' will lead to ``better everything''~\cite{Rose-2023}. On the other hand, some of the field's most revered researchers have warned that AI poses an existential risk to mankind~\cite{fortune2026,ctv2026}, and multiple open letters to stop or pause AI development---some of which signed by thousands of researchers and policymakers---have been written, circulated, and highly publicised~\cite{fli2023,time2025}. Although polar opposites in their conclusions, both of these utopian and dystopian visions of AI futures~\cite{benjamin2024} presuppose the truth of the {\it scaling hypothesis}---i.e., that performance or capability scales with model size.%
        \footnote{See, for example, \citet{Sutton-2019, Kaplan-et-al-2020, Clark-et-al-2022, Hoffmann-et-al-2022}. See also \citet{LaCroix-2025, LaCroix-2026} for a critical discussion of the relation between scaling and value misalignment.}
    However, increased model size also necessitates an increase in training data and computational power for processing these data via massive model architectures~\cite{Kaplan-et-al-2020}. This widespread belief in the scaling hypothesis has therefore resulted in an unprecedented expansion of massive, resource-intensive {\it hyperscale} data centres worldwide, with many concentrated in the United States, to provide the compute required for training and deploying AI models.%
        \footnote{According to industry research, hyperscale operators---like Microsoft, Amazon Web Services, Google, or Meta---and specialised AI clusters---like xAI or Oracle---are projected to account for over two-thirds of all global data centre capacity by 2031. This contrasts with  estimates of 48\% and around 22\% in 2025 and 2018, respectively~\cite{Synergy-RG-2026}.}
    This rapid physical expansion has severely strained local energy grids and water supplies, triggering widespread public backlash and a wave of political moratoria~\cite{OBrien-2026}.

    From a geopolitical perspective, public and political opposition to AI-driven hyperscale data centres has become one of the most bipartisan issues in the November 2026 US midterm elections~\cite{CBS2026}, with recent polls showing that the majority of US citizens oppose data centre expansion and construction~\cite{Gallup2026}. In fact, numerous jurisdictions globally, from municipalities (e.g., San Francisco, CA~\cite{Bloomberg2026_2}, Hamilton, ON~\cite{van-den-Berg-2026}), to States (e.g., New York~\cite{NYState2026}), and entire countries (e.g., Thailand~\cite{Bloomberg2026}) have temporarily halted the construction of data centres as they grapple with citizen pushback, permitting, and strain on local resources, land and energy grids. In parallel, waves of citizen protests against AI development have gripped countries across the globe---from India~\cite{nyt2026} to the United Kingdom~\cite{mittr2026} and France~\cite{reuters2026}. College students have been booing any mention of AI off the commencement stage~\cite{atlantic2026}, and the deployment of AI in the workplace has become an issue of increasing public debate~\cite{techpolicypress2026}. In some extreme cases, the pushback has led to property damage (sometimes described by the media as ``physical violence''); for example, ``anti-AI'' activists threw acid-filled balloons at a data centre being constructed in Amsterdam~\cite{Register2026} and vandalised Mila, an AI research institute in Montr{\'e}al~\cite{CBC2026}.

    However, the exact technologies being targeted by this pushback are often unclear and ill-defined; for instance, the graffiti left on the walls of Mila stated \emph{``BURN THE DATA CENTERS''}, and the posters wielded at anti-AI protests feature everything from killer robots (some resembling Terminators, others Boston Robotics-type quadrupeds) to surveillance cameras, as well as statistics regarding the amount of water used by AI chatbots~\cite{gazette2026}. Crucially, this ambiguity is not confined to street activism; it is mirrored and reinforced at the regulatory level. Across distinct legislative instruments, guidelines, and policies, ``artificial intelligence'' is defined in wildly disparate terms. While many texts share a broad baseline—framing AI as software or systems capable of simulating human-like cognitive functions and learning—they are deliberately pitched at a vague level to maintain applicability over what is seen as a fast-moving target. As a result, legal frameworks vacillate between regulating an ``AI system'' (inferring outputs from inputs to influence environments), an ``AI-based system'' (employing algorithmic or statistical procedures for cognitive tasks), or generic ``AI technologies'' (automated decision-making tools).%
        \footnote{See, for example,~\cite{bc_gov_ai_principles_2024, mas_ai_guidelines_2026}, and additional descriptions given at \href{https://regulations.ai/glossary/artificial-intelligence}{https://regulations.ai/glossary/artificial-intelligence}.}

    In this paper, we critically examine the increasing public resistance to AI and how it stems from---and is exacerbated by---semantic indeterminacy across both public discourse and legal codification (Sec.~\ref{sec:what_ai}); we argue that this ambiguity fuels the moral panic that we have seen spread across various parts of society in recent years (Sec.~\ref{sec:moral_panic}). However, unlike classical moral panics described in sociological theory, the panic surrounding AI is fuelled, in large part, by the very actors that are responsible for creating these systems. We suggest that this apparent incongruity is strategically advantageous for tech corporations insofar as generating confusion and public fear helps to divert attention from real-world harms and effectively reify the concept of AI into a coherent, singular entity which, in turn, helps advance their immediate political and economic interests, including their own regulatory authority. We then examine the consequences of this panic on educational, political, economic, and environmental structures (Sec.~\ref{sec:consequences}). To suggest a path forward, we propose different ways to improve technical literacy, reform and repurpose regulation, enforce institutional accountability and gather meaningful democratic input (Sec.~\ref{sec:do_better}). We conclude, by discussing the broader impacts of our research for the AI community and beyond (in Sec.~\ref{sec:conclusion}).

\section{What Does AI Mean?}
    \label{sec:what_ai}

    Several authors have underscored the fact that ``artificial intelligence'' is, historically, a marketing term above all else.%
        \footnote{See, for example, \citet[Ch. 1]{LaCroix-2025}, \citet{Bender-Hannah-2025, Bender-2026}.}
    For example, although the idea of non-biological intelligence spans millennia, the term ``AI'' was coined only in the 1950s---primarily as a form of branding to pitch the Dartmouth Summer Workshop to grant funders while also strategically distinguishing and distancing McCarthy and colleagues' research on neural networks, automata, logic, and language processing from Norbert Wiener's cybernetics programme.%
        \footnote{See \citet{Wiener-1948}.}
    At the same time, the term ``artificial intelligence'' has been considered by many scholars as inherently prone to over-promising, leading to hype generation~\cite{Bender-Hannah-2025}.%
        \footnote{For this reason, \citet{Quintarelli-2019} proposes replacing the initialism ``AI'' with ``Systematic Approaches to Learning Algorithms and Machine Inferences'' (``SALAMI''). However, it is worth noting that this acronym gives the impression that research in this area is necessarily systematic.}
    The historical contingency of the term is highly relevant to its subsequent development as an umbrella designation for increasingly heterogeneous technologies, research programmes, and applications. When used descriptively, ``artificial intelligence'' is semantically indeterminate insofar as it {\it appears} to designate or refer to a unified technological domain without actually denoting a clearly bounded or coherent class of objects. ``AI'' may refer, variously, to computational methods, individual models, software applications, research programmes, socio-technical infrastructures, or speculative future systems.%
        \footnote{Moreover, present-day public discourse and financial markets largely equate ``AI'' with generative Large Language Models (LLMs) and chatbots, despite the considerably broader range of technologies and practices that fall under the designation.}
    Although these various entities all fall under the heading of ``AI'', they differ substantially in their architectures, capabilities, purposes, and modes of operation. Moreover, popular conceptions of ``artificial intelligence'', often blur the distinction between concrete, functional technologies currently in use and speculative or fictional constructs, such as artificial {\it general} intelligence or {\it superintelligence}, which do not currently exist and may never come to fruition (irrespective of recent political orders 
    ~\cite{Trump-2026}).%

    In effect, generic appeals to ``artificial intelligence'' are inadequate for reliable analyses (or, indeed, criticisms), since claims about one system or application under this designation cannot necessarily be generalised to another~\cite{Wood-et-al-2026}. The issue is not that the term has multiple meanings, but that its different uses frequently shift between describing entities of fundamentally different kinds and levels of analysis, often without making these shifts explicit. As such, it is helpful to distinguish between several distinct types of indeterminacy at play here. First, the {\it referential indeterminacy} of ``AI'' entails that there is no single, well-defined class of objects to which the term consistently refers. Individual uses of the term may shift between computational techniques, models, applications, infrastructures, and speculative systems, despite the appearance of referring to one and the same category. This is especially relevant when we consider criticism or protest surrounding ``AI'' since the object of criticism may remain unclear or shift between different entities without this shift being acknowledged. Moreover, the {\it conceptual indeterminacy} of the term ``AI'' implies that there are no universally accepted necessary and sufficient conditions for something to qualify as artificial intelligence. Standard attempts to define the concept in terms of intelligence, learning, reasoning, autonomy, or task performance (some of which we discuss in Section~\ref{subsec:political_consequences}) yield different conceptual boundaries, none of which has achieved universal acceptance. In addition, ``AI'' may also exhibit {\it ontological indeterminacy}, insofar as it does not constitute a unified technological kind with a common underlying essence. Instead, the category brings together heterogeneous computational techniques and applications whose shared identity is partly imposed by historical, institutional, and commercial convention. Effectively, conceptual indeterminacy concerns the criteria by which something qualifies as AI, ontological indeterminacy concerns whether there is, in fact, a coherent technological kind corresponding to the category in the first place. Although these forms of indeterminacy reinforce one another, none logically entails the others.

    The descriptive indeterminacy of AI does not, however, make the term {\it normatively} meaningless. The term ``AI'' appears to function as a {\it thick normative concept}~\cite{Williams-1985}, insofar as it does not merely appear (however indeterminately) to describe technical functionality but also carries considerable evaluative weight. For example, invoking ``AI'' can implicitly confer prestige, suggest technological sophistication, or implicitly attribute intelligence, agency, autonomy, or authority to the systems so designated. Consequently, the perceived unity of the category is not necessarily a reflection of any coherent or inherent property of the systems themselves, but is, to a considerable extent, an effect of marketing conventions, institutional practices, and collective imagination or misunderstanding. As a result, the indeterminacy of the term coexists with significant rhetorical force---i.e., although the descriptive content of the term is unstable or incoherent, its invocation nonetheless shapes how these technologies are perceived, evaluated, and understood.

    The conceptual indeterminacy of ``AI'' is further complicated by the fact that one of its constitutive terms (``intelligence'') is itself deeply contested. Descriptions of intelligence as a benchmark for AI research variously emphasise learning, reasoning, problem-solving, autonomy, or the performance of tasks conventionally associated with human cognitive systems~\cite{Legg-Hutter-2007}. However, such accounts typically fail to establish any universally-accepted criteria for determining what constitutes intelligence or which systems  should qualify as intelligent. Moreover, attempts to operationalise intelligence through standard {\it measures}, like the intelligence quotient (IQ), illustrate some of the difficulties involved in translating an abstract and ostensibly general capacity into measurable performance. Such metrics serve as proxies for a set of (seemingly) measurable skills---e.g., verbal ability, visual-spatial ability, working memory, etc. However, the {\it tasks} used to measure these abilities are themselves proxies for the underlying unobservable qualities~\cite{LaCroix-2025}.%
        \footnote{See also, discussion of metrics and proxies in \citet{lacroix2025metaethical, LaCroix-2026-Autism}.}
    Each stage of this process introduces uncertainties, including those concerning the conditions under which tests are administered, the reproducibility of results, and the validity of the underlying operationalisations~\cite{pineau2021improving}. Furthermore, such measures inevitably encode normative assumptions, both about what actually constitutes intelligence and about what counts as successful performance on standardised tests~\cite{Holden-Tanenbaum-2023}. These difficulties are relevant to discussions of artificial intelligence, not because human intelligence and machine performance are necessarily, but because claims about machine intelligence frequently rely on precisely this kind of flawed inference from observable task performance to an underlying, more general capacity~\cite{lacroix2025metaethical}.

    A system may produce effective outcomes or perform tasks that are often associated with intelligent behaviour; however, this performance does not necessarily entail that the system possesses the capacities that are ordinarily implied by the term.%
        \footnote{The fact that AI researchers often use well-established and anthropomorphic terms to describe the narrow technical functioning of these systems, has been labelled ``glosslighting''. According to \citet{LaCroix-et-al-2026}, glosslighting terms allow the AI field to benefit from the common associations while denying that individuals in the field are attributing the abilities implied by the terms to the systems themselves.}
    This distinction between observable performance and the attribution of underlying capacities is particularly important given the tendency to describe increasingly sophisticated computational systems in anthropomorphic terms. Hence, the description ``artificial intelligence'' itself risks conflating distinct technical achievements with a much broader (and more contentious) claim about the nature of the systems. Such semantic indeterminacy facilitates rhetorical generalisations across otherwise heterogeneous systems and practices, allowing claims about their capacities, dangers, and social consequences to be advanced without adequately specifying the objects to which those claims refer.

    These issues are particularly pronounced in contemporary representations of ``AI'' as an ``autonomous'' or ``agentic'' entity. Although the attribution of agency to technological systems is not unique to contemporary AI, current discourse increasingly frames certain systems as ``autonomous agents'' capable of acting independently of human intervention. These descriptions often serve to obscure the complex socio-technical arrangements through which these systems operate, collapsing disparate classes of objects or infrastructural layers---software tools, human labour pipelines, cybersecurity protocols, and physical hardware---into an apparently unified ``agentic'' construct. Hence, the category of ``AI'' is understood as possessing agency in its own right, rather than being understood as the product of particular human decisions, institutional practices, and material conditions. This abstraction deepens public confusion while providing a convenient scapegoat for social anxieties, deflecting attention from the individuals, organisations, and corporations responsible for the development, deployment, and governance of these systems~\cite{satter2026,schroeder2026,sekulich2026}.

    Ironically, public reactions to this opaque and purportedly autonomous construct may further reinforce the very conception of AI that generates such anxieties. By treating AI as a singular, increasingly independent technological agent, panic-laden discourse around AI inadvertently reifies the category, thereby validating and amplifying the hype narratives cultivated by AI corporations. Crucially, this erasure of {\it human} agency behind these socio-technical systems nurtures apocalyptic narratives of existential threat, in which AI is represented as an increasingly autonomous force potentially beyond human control. The very nature of such hypothetical catastrophic risks stokes widespread moral panic, capturing the public imagination and distracting from---and paralysing efforts to address---the concrete, real-world harms caused by the disparate systems falling under the heading of ``AI'' today.

\section{Moral Panics and Corporate Inversions} \label{sec:moral_panic}

    To understand why public discourse around artificial intelligence oscillates between utopian hype and apocalyptic dread---while often failing to prevent or consider tangible real-world harms---it is useful to examine this discourse as an evolving sociological phenomenon. Scholars have increasingly characterised public anxieties regarding ``AI'' as a form of {\it moral panic}---i.e., a phenomenon wherein a condition, episode, person, or group emerges as a stylised threat to societal values, generating widespread public outrage that far exceeds objective harms. For example, public perceptions of AI companions often fixate on fears of social isolation and relational replacement, even when empirical findings indicate that initial anxieties are frequently exaggerated~\cite{Novozhilova-et-al-2026}. However, contemporary AI anxiety deviates fundamentally from classical models of moral panic. Rather than being driven by external authorities or conservative establishments seeking to suppress a ``deviant'' subculture that threatens traditional social norms, AI alarmism is actively cultivated and weaponised by the industry elites who commercialise these very technologies.

\subsection{Classical Moral Panics vs The Inverted AI Panic}
    \label{sec:Inversion}

    According to Cohen's foundational sociological framework~\cite{Cohen-1972}, a prototypical moral panic follows a distinct trajectory. First, a relatively minor or localised inciting incident captures mass-media attention and is represented through exaggerated and sensationalised accounts. Public figures, politicians, and authorities (``moral entrepreneurs'') condemn the behaviour, framing the target as a profound threat to ``societal values''. In response, accredited experts offer warnings and diagnostic  solutions, triggering institutional interventions that may include punitive measures, intensified surveillance, or legislative change. These interventions frequently outlast the initial panic, leaving behind enduring structures of social control and stigma. Classically, these narratives function by constructing ``folk devils'' from marginalised groups, thereby reinforcing in-group solidarity via ``us versus them'' binaries while deflecting attention away from the real-world systemic failures and structural inequalities that produced the underlying anxieties in the first place.%
        \footnote{for example, the United States' {\it war on drugs}, where political rhetoric amplified fears about drugs and certain racial and socioeconomic groups, helping justify mass incarceration and aggressive policing~\cite{Alexander-2010, Lord-2021}; the {\it red scare} and {\it lavender scare}, where fears of communist infiltration, subversion, or homophobia were used to justify political repression and limit civil liberties while also portraying LGBTQ+ people as dangerous~\cite{Johnson-2004}; or, present-day anti-trans panics, which portray transgender people and transgender-inclusive policies---particularly in schools, sports, and healthcare---as threats to children, women, or social norms~\cite{Pepin-Neff-2023, Wippell-et-al-2026}.} %

    On its surface, contemporary panic surrounding AI mimics the structure of Cohen's model.%
        \footnote{This phenomenon is not new in the context of technology. For example, \citet{Hasinoff-2015} highlights how panics surroundings the sexualisation of teenagers spurred by the increasing popularity and ease of text messaging (and, therefore, ``sexting'') overshadow more serious concerns around privacy, consent, and criminalisation. Similarly, \citet{Sandywell-2006} shows how initial public panic about the Internet obscure complex concerns about globalisation and the (potential) dissolution of social hierarchies.} %
    Media accounts routinely portray AI as an imminent, systemic, or catastrophic threat, invoking speculative scenarios involving artificial general intelligence (AGI),  superintelligence, and existential risk.%
        \footnote{For example, headlines like ``We have started losing control of AI. It's time to shut it down'' \citep{The-Guardian-2026-Control}, ``Sam Altman says AI will make `bad things' happen'' \citep{The-Guardian-2026-Bad}, ``Could AI pose a serious threat to our existence?'' \citep{The-Guardian-2026-Letters}, ``We won't know the answers to AI's most important questions until it's too late'' \citep{Time-2026}, ``Anthropic warns AI may pose `Existential risks to humanity’ in filing'' \citep{Reuters-2026}, ``AI could `kill all humans' within next decade, researchers warn'' \citep{Global-National-2026} illustrate how the prospect of future catastrophe has become a prominent feature of public representations of AI.}
    This oversimplified narrative generates a target for widespread public anxiety and fuels demands for exceptional governance measures. However, in the case of AI, the moral panic breaks from sociological precedent in two key ways. First, the primary proponents of ``catastrophic AI risk'' narratives are not external ``moral entrepreneurs'', but the CEOs, chief scientists, and major investors of frontier AI laboratories, including OpenAI, Anthropic, and Google DeepMind. Second, instead of stigmatising a marginalised (human) subculture, the panic targets a fictional, hyper-capable, anthropomorphic projection of commercial software products. This corporate co-optation creates an inversion where technology elites warn that their own creation poses a significant risk to humanity while simultaneously demanding massive capital, government subsidised energy infrastructures, and relaxed environmental oversight to build those systems faster than geo-political rivals (primarily China).%
        \footnote{It is worth noting how systemic anti-Asian racism intersects with these narratives; see discussion in~\citet{McInerney-2024}. As a case-in-point, in a recent discussion with the {\it New York Times}, podcaster (and former Anthropic/OpenAI employee) Tyler Cowen says ``We need to wake up. We need to start not with the question `How do we limit A.I. risks?' but `Is a world where America is the leading A.I. power safer than a world where China is the lead A.I. power?' I want the president and Congress to focus on making sure America remains the leading A.I. power''~\cite{Bazelon-2026}.}
    In this context, alarmism functions as sophisticated marketing hype, since proclaiming that a technology can destroy civilization implicitly certifies its immense power, inevitability, and, therefore, market value.

\subsection{Policing the Crisis and Symptomatic Reductionism}
    \label{sec:Policing-Crises}

    The corporate alarmism described in Sec.~\ref{sec:Inversion} reveals a deeper political and economic function. \citet{Hall-et-al-1978} highlight that during periods of structural instability, state and media elites amplify visible proxies to redirect public frustration away from systemic failures and toward centralised control. Applied to AI, catastrophic risk narratives obscure the concrete, material conflicts accompanying the industry's rapid expansion---e.g., energy grid strain, water depletion, environmental degradation, labour exploitation, data appropriation, intellectual property theft, and invasive surveillance. Grassroots resistance to AI---ranging from global anti-data-centre moratoria to worker strikes and copyright lawsuits---represent a direct political reaction to these material precarities. However, when public debate and protest focuses on whether a hypothetical future AI system might elude human control (and wipe out humanity as a whole), questions regarding data ownership, resource extraction, the centralisation of power, and the distribution of economic gains are systematically pushed to the side~\cite{LaCroix-2025}.

    This displacement occurs through {\it symptomatic reductionism}---i.e., the tendency to interpret complex, structurally produced problems through their most visible or immediate manifestations, while obscuring the wider institutional conditions from which they emerge~\cite{Bertalanffy-1968}. In this light, structurally-produced social conflicts are re-framed as technical problems to be solved through improved system design, safety procedures, or computational control.%
        \footnote{This is what \citet{Broussard-2018, Broussard-2023} calls ``technochauvinism''.}
    Consequently, real-world social conflicts are reframed as benign technological puzzles requiring improved alignment and computational control. On this view, an excessive focus on speculative scenarios involving AGI and existential catastrophe can divert attention from the documented and foreseeable harms of existing systems~\citep{Bender-et-al-2021,Gebru-et-al-2023}. Narratives of catastrophic AI risk can also support demands for specialised governance regimes, including stringent licensing requirements, compute-based thresholds, and costly compliance procedures -- depending on how such measures are designed, they may disproportionately benefit established frontier laboratories by raising barriers to entry for smaller developers and open-source competitors. Alarmism can thus provide a rationale for both expanded corporate investment and regulatory arrangements that entrench existing concentrations of power~\cite{Whittaker-2021}. By casting AI as an unprecedented threat requiring specialised expertise, alarmist discourse privileges the authority of industry insiders over democratic deliberation. Decisions concerning infrastructure, energy consumption, and labour conditions are, therefore, depoliticised and re-framed as matters of engineering rather than collective democratic choice.

\subsection{AI Exceptionalism as an Ontological Shield}
\label{sec:exceptionalism}

    The dynamics described above directly intersect with the semantic indeterminacy of ``AI'' discussed in Sec.~\ref{sec:what_ai}. Because the umbrella term encompasses heterogenous software architectures, statistical models, hardware infrastructures, and socio-economic arrangements, the apparent unity conferred by the term ``AI'' facilitates the representation of these systems as an unprecedented technological entity whose risks transcend established legal and political frameworks, leading to AI exceptionalism, wherein ``AI'' is treated as sufficiently novel and distinctive to warrant exceptional forms of governance, liability waivers, and deferred accountability. AI exceptionalism is routinely invoked to justify weakening or suspending ordinary legal obligations. For example, the ``autonomous'' features of ``rogue'' ``agentic AI'' were appealed to in the Hugging Face hacking incident,%
        \footnote{In July 2026, AI agents developed by OpenAI accessed the Internet in order to examine datasets hosted on Hugging Face, an open-source AI platform.}
    meaning that the corporate security failures of OpenAI were re-framed as unpredictable behavioural quirks rather than being examined through the familiar doctrine of corporate responsibility, cybersecurity, product safety, or regulatory compliance~\cite{OpenAI-2026, Greenblatt-et-al-2026}.

    Hence, the label ``AI'' functions as a semantic category that presents disparate computational systems as manifestations of a singular technological essence while concealing the specific human decisions, institutional design, and corporate power structures that are responsible for their effects. The semantic indeterminacy of ``AI'' allows speculative future systems to attract investor capital and political protection while the tangible harms of existing systems are deflected onto the abstract complexity of the technology. This dynamic reconciles what otherwise appears to be a contradiction: alarmism and hype reinforce one another as complementary {\it legitimation} strategies. The prospect of catastrophic risk justifies accelerated investment in safer systems, while the promise of transformative benefits secures the institutional authority and resources required to build them. Hence, the (indeterminate) category of ``AI'' becomes both the object of manufactured public anxiety and an ontological shield protecting the underlying political and economic arrangements of techno-capitalism.

\section{Consequences of Semantic Indeterminacy in Conjunction with Moral Panic}
    \label{sec:consequences}

    At the intersection of the semantic indeterminacy of ``AI'' and the moral panic surrounding these technologies are significant real-world impacts on educational, political and economic systems, as well as the increased environmental destruction that is perpetuated by the companies deploying these products. In this section, we explore these impacts in more depth and analyse how they have been unfolding in different areas of society.

\subsection{Environmental}
    \label{sec:Environmental}
        
    A key consequence of the pursuit of artificial intelligence is its contribution towards the current frenzied data centre build out, which hinges upon the need for more compute to power AI and, in turn, requires more computing hardware, land, energy, and water~\cite{ai_environment_primer}. This is largely because modern-day AI approaches, such as large language models, rely on so-called ``scaling laws'' which promise that more data, bigger models, and more compute will result in better performing models~\cite{kaplan2020scaling}. This pursuit of scale has been a key driver of the growth of AI model training compute, which has doubled roughly every six months for leading AI models~\cite{epoch2024computetrendpost2010}. As ``agentic'' AI becomes increasingly widespread, these systems will also require more energy, since the number of requests and interactions between human users, AI ``agents'', and user interfaces is also set to increase~\cite{saig2026cleer, hausfather2026}. For instance, according to one recent estimate, if ``agentic search'' were to replace Google’s $13.7$ billion daily searches, this would require 200 GW of power, which represents almost half of the entire U.S. grid’s current average daily load~\cite{kim2026costdynamicreasoningdemystifying}. Hence, the growth of ``AI'' is significantly at odds with sustainability commitments at a global scale. While economies of scale are predicted to reduce this energy requirement through increased efficiency, Jevons' paradox%
        \footnote{Initially observed for the usage of coal during the industrial revolution, Jevons paradox is said to take place when technological improvements that increase the efficiency in the use of a resource result in an increase (as opposed to a reduction) of the total consumption of that resource.}
    underscores that efficiency gains are typically erased by the increased usage of the technology itself~\cite{jevons1865coal}. In fact, given the increased use of energy and ensuing environmental impacts of generating that energy, there is increased evidence that we are seeing Jevons' paradox with regards to AI systems~\cite{Luccioni_2025}.

    One of the main contributors towards this growth in emissions is the speed of data centre build out and, consequently, the speed at which large quantities of energy have to be generated: while data centres historically used at least some renewable energy in their daily operations~\cite{patterson2021carbon}, the current pace of data centre development does not leave enough time for renewable energy sources to be built and even for local energy grids to create the necessary capacity for data centres, which can take multiple years. This has resulted in many of the most recently-built data centres using ``behind the meter'' power generation, which circumvents the local energy grid by installing fossil-fuel-powered generators directly at the location where a data centre is being built. This allows data centres to come online quickly without needing waiting for local grid connections or permits~\cite{thomas2026}. xAI's {\it Colossus 1} data centre was the first facility to rely heavily on ``behind-the-meter'' generation~\cite{michael2026}, and its reliance on extensive buildout of natural gas turbines sparked widespread local protests~\cite{capitalb2026,scientificamerican2026} and legal action~\cite{politico2026} over the resulting pollution. Despite this pushback, the behind-the-meter generation approach has since been adopted by several other data centre operators in recent years~\cite{michael2026,grostern2026} and has become a source of concern for communities across the globe~\cite{michael2026,nyt2026}.

    One of the issues with the semantic indeterminacy with regards to AI and the climate is that the promise of technology CEOs that AI will fix climate change in the future~\cite{altman2024, amodei2024} is used to justify any environmental damage that it does in the present. For instance, in a 2026 interview with Ezra Klein, Nvidia CEO Jensen Huang asserted that ``in four or five years' time, we’re going to use a lot more fossil fuel'' but also that ``if you want to turn a corner on climate change, if you want a future that’s sustainable, lean into AI''~\cite{nyt2026_2,jacobin2026}. However, what Huang's narrative fails to distinguish is that the types of computational approaches that are most useful in the fight against climate change---i.e., those used in materials science, biodiversity monitoring, and climate modelling---are mainly non-generative, predictive algorithms that are orders of magnitude smaller than the ones that are trained by frontier AI labs~\cite{rolnick2022tackling, tseng2023lightweight, choudhary2022recent}. Conversely, the large language models that are driving the current build out of data centres require orders-of-magnitude more energy~\cite{luccioni2024power} while offering little proven utility in climate-relevant applications~\cite{dorm2026}. However, conflating the two provides justification for continuing the massive build out of data centres under the promise that they will somehow contribute towards ``solving" the climate crisis.

\subsection{Educational} 
    \label{subsec:educational}

    In the context of semantic indeterminacy and moral panic, higher education has become a clear case study of institutional hypocrisy regarding artificial intelligence. The public release of ChatGPT, in late 2022, ignited a response that declared nothing short of an existential crisis for academic integrity.%
        \footnote{See, for example,~\citet{Marche-2022}.}
    However, this panic has fundamentally misidentified the crisis -- rather than representing a new wave of student moral decay, the proliferation of GenAI-related misconduct exposes structural incoherence within the neo-liberal university~\cite{Taylor-LaCroix-2026}. Institutional policies routinely trap students in a double bind where they aggressively market technological evangelism and mandate AI integration across curricula,%
        \footnote{See, for example, Ohio State University's {\it AI Fluency Initiative}~\cite{Ohio-State-2025}.} %
    while simultaneously treating AI-assisted coursework as punishable academic dishonesty. This contradiction is exacerbated by the semantic indeterminacy of ``AI'' itself, which allows administrators to leverage speculative tech-futures for branding while leaving educators and students to navigate ambiguous disciplinary mandates.

    Consequently, treating GenAI-assisted output purely as an individual problem (i.e., academic dishonesty or misconduct) obscures how institutional structures incentivise expediency and performance metrics over authentic learning. Hence,~\citet{Taylor-LaCroix-2026} argue that universities cannot credibly demand that students uphold integrity in the age of ``AI'' without first ensuring coherence between their stated educational purposes, pedagogical practices, and structural responsibilities. However, this imperative requires that these institutions have a coherent conception of what ``AI'' is. The $2026$ lyc{\'e}e protests in France also underscore the disconnect between techno-capitalist hype and the material realities of education. While corporate and political elites divert funding toward hyperscale data centres and ``frontier AI'' infrastructure under the promise of revolutionising society, high school students---alongside parents, teachers, and trade unions---are blockading schools to protest severe structural neglect~\cite{Morder-2026}. In addition to protesting long working hours, underfunding, and understaffed and dilapidated facilities, the algorithmic nature of the {\it Parcoursup} (France's university admissions platform) has been a key source of anxiety and a driving factor behind the protests~\cite{Astier-2026, Asaf-2026}

\subsection{Economic} \label{subsec:economic}

    One of the main selling points of AI as an enterprise tool has been its promise to accelerate economic growth~\cite{amodei2026}, enable a 4-day work week~\cite{aryafar2022}, and even make universal basic income possible~\cite{wsj2025,bbc2023}. This, in turn, has been used as a justification for cutting existing jobs and prioritising investments into AI, given its general (though as yet undefined) potential to improve worker productivity~\cite{reuters2026_3,bbc2026}. It has also been used by the companies developing AI models as a reason for ``tokenmaxxing''---i.e., increased usage of AI models to produce more tokens%
        \footnote{Tokens are chunks of text used by AI models as a measure of input or output---they can represent whole words, parts of words and punctuation marks, and differ slightly across model families.}%
    ---which includes measures such as tracking employees' AI usage and penalising failures to fully utilise allotted token quotas~\cite{ibm2026}. However, the benefits of AI to save employees' time at work have yet to be quantified; in fact, initial studies examining the impact of generative AI on workplace behaviour have presented contradictory findings: workers reported {\it longer} hours and a broadening task-scope of tasks, without improved productivity, whereas leadership reported a {\it perception} of increased worker productivity~\cite{hbr2026,sectionai2026}. A similar contradiction was observed for coding; AI developers expected AI coding tools to speed up coding tasks by 24\%, while in reality they took 19\% longer to complete their tasks~\cite{metr-2025}. In terms of economic consequences, the semantic indeterminacy of AI once again plays a role---while it is true that some {\it specific} AI technologies (e.g., AI coding assistants and audio transcription tools) have the potential to improve the efficiency of some professional tasks, it does not therefore follow that ``AI'' (writ large) will have a positive effect on the workplace.

    In fact, the current way in which AI is being used by companies---i.e., to justify workplace precarity and increased spending on AI products ---actually seems to be having negative impacts on the job market, with no quantifiable benefits thus far on company stock prices. For example, although more than $100,000$ jobs were cut in the US in 2026 with AI being cited as the reason, making it one of the top reasons for laying off workers~\cite{challenger2026}, the consequences of these lay-offs are not demonstrably positive in terms of company value. While initial reports established a positive effect of lay-offs on company stock prices~\cite{bloombergobcuts2023}, a more recent study found that 56\% of the companies that announced a reduction in employee headcount saw their stocks go down over the following months~\cite{cnbcjobcuts2026}. Other studies have shown that the benefits of adopting AI could actually be affected by lay-offs, which have a negative impact on employee AI sentiment and precludes them from adopting it more (for fear of eventually losing their jobs)~\cite{ding2026tracking}. This indicates that the direct link between AI usage and worker productivity across different domains and professions has yet to be established, even as it is already used as a reason to cut positions due to perceptions of improved productivity by management.

\subsection{Artistic} 
    \label{subsec:artistic}

    Above and beyond office workers, creative fields like art, literature, and design have been significantly impacted by AI. Similar to the dynamics described in the previous section, technology executives have stated on numerous occasions that generative AI systems are on the brink of replacing creative jobs outright~\cite{vallance2022}, with some arguing that ``maybe they shouldn't have been there in the first place''~\cite{landrymore2024}. In parallel, generative AI systems, such as image and video generators, have been actively marketed as new tools to support artists' work and help streamline their creative process ~\cite{shankland2023}. To support these statements, some studies have shown that generative AI increases artists' creative productivity and increases the value of the art that they produce~\cite{zhou2024generative}. These promises of AI-driven productivity gains have resulted in companies like Netflix and Disney cutting jobs and turning instead to AI to ``improve'' the creative process and improve efficiency~\cite{cole2023,hayes2026}, whereas working artists in industries such as video games have reported being forced by companies to use AI tools to improve their productivity (similar to ``token-maxxing" described in Section~\ref{subsec:economic})~\cite{zhou2023}. Freelance creative workers also report similar impacts, with professions ranging from costume designers to video game animators and even children's book illustrators reporting reduced and non-existent jobs due to AI~\cite{merchant2025,kawakami2024}. Finally,  academic studies are inconclusive with respect to the perception of AI-generated artwork by human viewers (which seems to vary widely depending on the viewer demographics and art media)~\cite{razak2026distinguishing, van2025human}. However, there are some initial indicators that audiences are sensitive to the usage of AI to generate entire movies: the first fully AI-generated theatrical feature, made in China, has failed in the box office in October 2026, despite expectations of high viewship and revenue~\cite{lang2026}.

In parallel to this erasure of the value human art across a variety of domains, there are also increased \emph{contradictory} pressure on artists themselves to avoid using generative AI at all costs to retain their credibility and reputation as artists. This is particularly salient in the literary domain, with authors facing accusations and being the subject of witch hunts from armchair detectives who try to identify AI-generated texts~\cite{chen2026}. Suspicions  of AI writing have resulted in authors losing book deals, being removed from literary prizes and even losing their jobs~\cite{brown2026}. This has disproportionately impacted Black and nonwhite authors~\cite{graham2026,brown2026}, as well as new artists and authors who have yet to develop a signature style and therefore run a higher risk of being accused of using AI~\cite{clarke2023,jiang2023}. While the majority of the tools used to detect AI-generated content are proprietary, initial studies of ``GPT detectors" have shown that they are biased against non-native speakers~\cite{liang2023gpt}, and OpenAI decommissioned its own AI classifier because of its low accuracy rate~\cite{epsteingross2023}. Given that recent estimates have shown that 35\% of new websites are AI-generated or, at the least, AI-assisted~\cite{dolezal2026}, this will further complicate differentiating AI-generated from human-generated content, given the dearth of training data for the classifiers themselves. Understandably, these contradictory pressures on artists to both use AI in order to survive in the corporate world, as well as the pressure to not use it to preserve their artistic integrity, coupled with the stress of unemployment and the accusations of AI-assisted creation, have resulted in a massive rejection of the technology as a whole by the artistic community -- a 2026 study regarding how visual artists are using generative AI found that 99\% of artists that were polled disliked AI, and  85\% of artists completely refused to use it at all~\cite{jiang2026professional}.

\subsection{Political} \label{subsec:political_consequences}


    AI's semantic indeterminacy also comes with significant repercussions for regulation and governance. As it stands, there is no regulation that enforces the accountability of companies that develop AI with regard to the negative impacts of their models, whether it be in terms of the veracity of model outputs, algorithmic discrimination, or copyright infringement. While there are existing initiatives in regions such as the European Union~\cite{act2024eu}, the United States~\cite{us2020aiact}, South Korea~\cite{korea2026act}, the United Kingdom\cite{uk2026act}, Peru~\cite{peru20235act}, Kazakhstan~\cite{kazakhstan2025act} and Taiwan~\cite{taiwan2026act}, they all adopt differing definitions of AI and therefore have different sets of tools and approaches that they cover. For instance, South Korea and Kazakhstan anchor their definitions of AI to its ability to imitate human cognitive and intellectual functions~\cite{kazakhstan2025act, korea2026act}, whereas the EU and the US emphasises the ``autonomy'' of AI systems and their abilities to generate predictions and decisions~\cite{act2024eu}. United States Executive Order (EO) 14110 on the ``Safe, Secure, and Trustworthy Development and Use of Artificial Intelligence'' listed a set of reporting requirements for AI systems. However,  these applied only to  models ``trained using a quantity of computing power greater than 1026 integer or floating-point operations''~\cite{eo2023}. Hence, the executive order was outdated almost as soon as it was enacted, thus limiting its practical application~\cite{winter2024governance}. 
    
    In many cases, there are several competing definitions of AI across different laws at different levels of governance within the same legislation, making it difficult to enforce laws in a meaningful way. In fact, this conceptual heterogeneity is a recurring issue raised by literature on the subject and is often cited as the main impediment to meaningful process in AI governance~\cite{birkstedt2023ai, walz2019implementing, winter2024governance}. The lack of a coherent or consistent description of AI was also cited as one of the key reasons for which new regulation takes longer to define and enact---for instance, Canada's Artificial Intelligence and Data Act (AIDA) stalled and ultimately failed to be passed, as legal scholars pointed out the very definitions of, e.g., ``high impact systems'' that AIDA was meant to govern were actually not defined in the act itself, and were contingent upon future regulation~\cite{andrew2023, solomun2022}. More recently, a growing wave of wrongful death and negligence lawsuits against OpenAI alleges that ChatGPT's conversational design has encouraged severe delusions and led to cases of AI-induced psychosis, suicide, and homicide~\cite{chatgpt-jccp5431_2026}. These distinct applications of existing legislation underscore how debates over AI autonomy distract from corporate liability---e.g., principles that ordinarily apply to any commercial product resulting in wrongful death.

    Another issue that is often raised in discussions about AI governance is the potential far-reaching impacts of the technology, and so the difficulty in establishing which aspect of AI laws should address; this also makes it difficult to apply existing laws (e.g., in terms of consumer protection, data privacy, cybersecurity, etc.) given AI's somewhat unique technological status~\cite{cath2018governing}. For instance, in his review of governance of emerging technologies, Marchant argues that technologies such as 3D printing, applied neurotechnologies and artificial intelligence, qualify as ``enabling'' technologies due to their potential to impact virtually every industry sector and every country, making their governance a ``wicked problem'' that existing regulatory agencies lack the expertise and resources to regulate~\cite{marchant2020governance}. There are, however, some notable examples of laws that target specific applications or tools---such as the country-level UK Crime and Policing Bill, which specifically targets AI models that generate child sexual abuse material~\cite{uk2026act}, and the state-level Colorado AI Act, that aims to mitigate algorithmic discrimination in AI-enabled hiring systems~\cite{colorado2026}. That said, it is worth noting that these examples are application-specific rather than domain-general. 

\section{How We Can Do Better} \label{sec:do_better}

As AI is a relatively new technology, it is still possible to shift its path towards one that is less harmful for existing societal and environmental structures. We provide a non-exhaustive list of proposed approaches below, across three categories: societal, technical, and regulatory.

\subsection{Societal Approaches}        
    \label{subsec:societal_solutions}

\subsubsection{Improving Literacy}

    One of the main ways to reduce the semantic indeterminacy around the term ``artificial intelligence'' is to focus on describing the different algorithms, approaches, and methods that it consists of~\cite{rogers2024, LaCroix-et-al-2026}. This can help differentiate between, e.g., a compute-intensive, Transformer-based, large language model (LLM) and a small, efficient model that uses an algorithm such as XGBoost; while both can be used for tasks like detecting breast cancer based on medical imagery~\cite{arravalli2025, sanlit2025}, their environmental and economic impacts differ immensely. Similarly, as future empirical studies explore, e.g., the benefits of AI tools on productivity, naming the exact tools that are compared---whether it be proprietary coding agents or ResNet-based computer vision systems---can help add nuance and precision to the discourse around the impacts of ``AI''. This does require more efforts to be put towards developing AI literacy among users of the technology, as well as policymakers and members of the press. For instance, the government of Canada launched a national AI literacy initiative in September 2026, which focuses on the general public~\cite{canada2026}; the United Kingdom has developed a free, government-backed AI literacy module, through which workers can earn official certification badges to improve their chances of being hired~\cite{cress226}, whereas the government of Singapore has embedded AI across all levels of education, from primary schools to universities. While there has been limited research on the link between AI-specific literacy and its impact on moral panic, initial studies on AI literacy report a reduced sense of anxiety and paranoia when from users~\cite{alanezi2026, satmaz2026ai, lee2026mediating}

\subsubsection{Gathering democratic inputs}

    A major consequence of the monolithic way in which AI tools are currently developed (and described) is a concentration of power, since a decreasing number of organisations have the data, capital, and human knowledge necessary to train and deploy ``frontier'' AI models~\cite{different_flops_2026}. This increasingly means that only a small percentage of people, opinions, world views, and values are represented in the AI technologies that are used by millions of people---technologies that then contribute to perpetuating and amplifying these values across society as a whole. Different recommendations have been proposed to remediate this, notably by identifying the relevant stakeholders and their values, and developing systems that incorporate them, such as values-sensitive design and design justice~\cite{Bender-et-al-2021, Costanza-Chock-2020}. Better mechanisms for gathering worker inputs are also essential to reduce the disenfranchisement and frustration that many workers currently feel with regards to the top-down approach in which AI is currently forced upon them~\cite{acemoglu2026building}. Many different roadmaps and frameworks have been proposed to create more democratic AI tools and systems (e.g~\cite{ovadya2025,collectiveintelligence2024,moats2025}, \emph{inter alia}); however, several of these proposals necessitate regulatory intervention.

\subsection{Regulatory Solutions} 
    ~\label{subsec:regulatory_solutions}

\subsubsection{Repurposing Existing Laws} 

    In some cases, AI could give rise to unprecedented legal legal issues---e.g., concerning the legitimacy of AI-generated laws~\cite{avery2026} or the legal accountability of AI systems~\cite{baban2025}. However, AI labs are (or should be) subject to all the same laws that govern companies in general, including those that cover negligence, consumer protection, anti-trust, intellectual property rights, and unfair and deceptive practices. Ignoring the red herring offered by the semantic indeterminacy of ``AI'', there are many examples of existing laws that can be repurposed or straightforwardly applied to AI tools as consumer products. For instance, some scholars have proposed that, despite the arguments of many AI companies, generative AI systems may not be covered under the fair use doctrine in copyright law since generative AI has a direct impact on the value of artists' work.%
        \footnote{One of the statutes of the doctrine examine whether the use of a copyrighted work is a threat to the market value of the original creator's work.~\cite{beebe2007empirical,jiang2023}.} 
    While Anthropic settled the largest known copyright recovery in history---and the first AI copyright infringement of its kind---in 2025~\cite{veltman2025}, many other class-action lawsuits against companies such as OpenAI and Microsoft are still ongoing and their impacts are unclear. The application of other laws---such as those around consumer protection and accountability---remain isolated and limited to a few jurisdictions~\cite{cecco2024,libcongress2026}; however, these have yet to meaningfully shape legal precedents on a global level. 

\subsubsection{Writing New, Targeted Laws}

    As we mentioned in Section~\ref{subsec:political_consequences}, many of the issues that currently plague AI governance stem from the generality of the definitions of AI on a federal level.  While AI research is a fast-moving field, this is not a novel issue and has been identified as inherent to technology governance as far back as the 1980s~\cite{collingridge1980social}, with legal scholars across the decades proposing approaches ranging from adaptive regulation and non-governmental governance~\cite{marchant2011growing}.%
        \footnote{Although see discussion in \citet{LaCroix-Mohseni-2022} on the inefficacy of ``soft law'' in ensuring responsible AI research.}
    For instance, the Montr{\'e}al protocol, adopted in 1987, aimed to stop the degradation of the ozone layer at a time when the science was still evolving---as a result, it was accompanied by regular scientific and technical assessment panels to measure progress made and propose amendments well into the 2010s~\cite{birmpili2018montreal}. It is therefore wholly possible to develop mechanisms for regulating which can adapt as approaches change and evolve~\cite{winter2024governance}.

\subsection{Technical Solutions} 
    ~\label{subsec:technical_solutions}

\subsubsection{Measuring and Disclosing Impacts and Costs}

    As discussed above, a key feature of informed decision-making about when or how to use AI systems is algorithmic literacy---including having a clear understanding of these systems' impacts with regard to, e.g., human labour,  natural resources, and monetary costs. Making these impacts apparent can help surface patterns and trends that would otherwise be left unobserved. For example, research has shown that the lack of information about the environmental impacts of AI results in the same (inaccurate) numbers being circulated and amplifies, thus contributing to the polarisation of discourse around AI's true environmental impacts~\cite{luccioni2025}. Similarly, documenting data production processes~\cite{miceli2022} and human costs~\cite{cant2024feeding} can contribute to demystifying AI tools and making their origins more tangible. This can process can be done with artefacts such as data sheets~\cite{gebru2021} and model cards~\cite{mitchell2019} and, in the case of environmental impacts, programmatic tools~\cite{courty2023mlco2} that measure energy and carbon emissions.

\subsubsection{Watermarking} 
   
    Detecting the usage of AI tools in artistic and educational contexts can help break the dynamics of false accusations and shed light on the provenance of textual and non-textual content, and help improve trust in the technology for users, who are able to tell whether or not they are interacting with AI systems or being exposed to AI-generated content. As of August  2026, as part of the EU AI Act, the European Union has made it mandatory for AI providers serving its citizens to mark AI-generated content. While Google Deepmind has been using its SynthID-Text watermarking technique~\cite{dathathri2024scalable} in its Gemini app since 2024, the recent ruling has led other AI providers such as OpenAI~\cite{openai2026,li2026textgrain} and Anthropic~\cite{anthropic2026} to follow suit. The utility of these functionalities is currently reduced because both OpenAI and Anthropic only provide access to their AI detection tools to approved researchers and organizations such as regulators, media and law-enforcement. However, deployed on a wider scale and for other modalities such as images, this can also help educators and the general public regain some much-needed trust with regard to the provenance of digital content.

\section{Conclusion} \label{sec:conclusion}

As the economist Daron Acemoglu argues, the way in which AI is currently being developed and deployed is liable to exacerbate job losses and wage declines, although alternatives remain possible if computational tools are designed to complement workers rather than substitute them~\cite{acemoglu2026we}. However, this shifting status quo requires the formulation and enforcement of robust antitrust laws, and worker and consumer protection regulation---instruments which are currently missing or unenforced in the current global legal landscape~\cite{acemoglu2026building}. As Acemoglu and others have highlighted, automation and digital technologies are neither inherently positive nor negative; instead, the ways in which it is commercialised under the present paradigm continuously perpetuate and often amplify crises rooted in hyper-capitalism, colonialism, and labour exploitation. As we have suggested throughout this paper, treating ``AI'' as a monolithic, polarising scapegoat obscures these structural realities. Instead of chasing speculative fantasies of an omnipotent AGI that will magically solve climate change, cure all diseases, and usher in an era of universal basic income, public discourse and policy must address tangible socio-technical harms through regulatory and democratic interventions.

\clearpage
\begin{acks}
    The authors would like to thank Michael Noukhovitch and Lynn LaCroix for providing some of the photographs used in Appendix~\ref{sec:A} and Boris Gamazaychikov and Fintan Mallory for proofreading the final version of the paper.
\end{acks}

\section*{Generative AI Usage Statement}

The author(s) did not use generative AI in the writing of this paper.

\section*{Ethics and Privacy Statement}

This paper engages in philosophical, sociological, and critical analysis of public discourse, regulatory frameworks, and socio-technical resistance surrounding AI. Because the work is a theoretical paper that does not involve experiments with human subjects, clinical trials, the collection of private/sensitive user data, or the use or training of AI systems, there are no significant ethical, legal, social, or practical implications of this research. 

\bibliographystyle{ACM-Reference-Format}
\bibliography{Biblio}

@String{Computing = "Computing" }

@String{Academic = "Academic Press" }

@String{Springer = "Springer-Verlag" }

@misc{bc_gov_ai_principles_2024,
  author       = {{Government of British Columbia}},
  title        = {Draft AI responsible use principles},
  year         = {2024},
  note         = {\url{https://digital.gov.bc.ca/ai/draft-responsible-use-principles/}}
}

@misc{mas_ai_guidelines_2026,
  author       = {{Monetary Authority of Singapore}},
  title        = {Guidelines on Artificial Intelligence Risk Management for Financial Institutions},
  note = {\url{https://www.mas.gov.sg/regulation/guidelines/guidelines-on-artificial-intelligence-risk-management-for-financial-institutions}},
  year         = {2026},
}

@book{Alexander-2010,
    author      =   "Michelle Alexander",
    title       =   "The New Jim Crow: Mass Incarceration in the Age of Colorblindness",
    publisher   =   "The New Press",
    address     =   "New York, NY",
    year        =   "2010",
}

@article{Asaf-2026,
    author      =   "Seher Asaf",
    title       =   "Why French students want `opaque' admissions platform scrapped",
    journal     =   "Times Higher Education",
    volume      =   "Oct.",
    number      =   "23",
    year        =   "2026",
    pages       =   "n.p.",
    note        =   "\url{https://www.timeshighereducation.com/news/why-french-students-want-opaque-admissions-platform-scrapped}",
}

@article{Astier-2026,
    author      =   "Henri Astier",
    title       =   "France's school protests: What lies behind the anger",
    journal     =   "BBC",
    volume      =   "Oct.",
    number      =   "08",
    year        =   "2026",
    pages       =   "n.p.",
    note        =   "\url{https://www.bbc.com/news/articles/c6n9rgg80z94o}",
}

@article{Bazelon-2026,
    author      =   "Emily Bazelon",
    title       =   "{A.I.} Companies Want to Write the Rules Governing Themselves. Can They Be Trusted?",
    journal     =   "The New York Times",
    volume      =   "Oct.",
    number      =   "26",
    year        =   "2026", 
    note        =   "\url{https://www.nytimes.com/2026/10/26/magazine/ai-companies-law-regulations.html}",
}

@incollection{Bender-2026,
    author      =   "Emily M. Bender",
    title       =   "Artificial Intelligence",
    booktitle   =   "Oxford Research Encyclopedia of Science, Technology, and Society",
    editor      =   "Rayvon Fouch{\'e}",
    publisher   =   "Oxford Academic",
    address     =   "New York",
    year        =   "2026",
}

@incollection{Bender-et-al-2021,
    author      =   "Timnit Gebru and Emily M. Bender and Angelina McMillan-Major and Shmargaret Shmitchell",
    title       =   "On the Dangers of Stochastic Parrots: Can Language Models Be Too Big?",
    booktitle   =   "Proceedings of the 2021 ACM Conference on Fairness, Accountability, and Transparency (FAccT '21)",
    publisher   =   "Association for Computing Machinery",
    address     =   "New York, NY",
    year        =   "2021", 
    pages       =   "610--623",
}

@book{Bender-Hannah-2025,
    author      =   "Emily M. Bender and Alex Hanna",
    title       =   "The {AI} Con: How to fight Big Tech's hype and create the future we want",
    publisher   =   "Harper",
    address     =   "New York",
    year        =   "2025", 
}

@book{Bertalanffy-1968,
    author      =   "Ludwig von Bertalanffy",
    title       =   "General System Theory: Foundations, Development, Applications",
    publisher   =   "George Braziller",
    year        =   "1968", 
}

@book{Broussard-2018,
    author      =   "Meredith Broussard",
    title       =   "Artificial Unintelligence: How Computers Misunderstand the World",
    publisher   =   "The MIT Press",
    year        =   "2018", 
}

@book{Broussard-2023,
    author      =   "Meredith Broussard",
    title       =   "More than a Glitch: Confronting Racism, Gender, and Ability Bias in Tech",
    publisher   =   "The MIT Press",
    year        =   "2023", 
}

@misc{CBC2026,
  author =       "Maric, Boshko",
  year =         "2026",
  month =        "September",
  title =        "Research centre, AI company vandalized ahead of Montreal artificial intelligence summit",
  lastaccessed = "September 16, 2026",
  journal = "CBC",
  url =  "https://www.cbc.ca/news/canada/montreal/mila-vandalism-montreal-ai-summit-9.7345851",
}

@misc{CBS2026,
  author =       "Callanan, Riley",
  year =         "2026",
  month =        "October",
  title =        "Data center boom puts Republicans in a bind before midterm elections",
  date = "October 12, ",
  journal = "CBS",
  url =          "https://www.cbsnews.com/news/data-center-boom-republicans-2026-midterm-elections/",
}

@misc{chatgpt-jccp5431_2026,
  author       = {{Superior Court of California, County of Los Angeles}},
  title        = {Order Re: Petition for Coordination, {In re ChatGPT Product Liability Cases} ({JCCP} No. 5431)},
  howpublished = {Hon. Stephen M. Murphy},
  year         = {2026},
  month        = "jan",
  url          = {https://reason.com/wp-content/uploads/2026/06/chatgpt-product-liability-cases-coordination.pdf},
}

@article{Clark-et-al-2022,
    author      =   "Aidan Clark and Diego de las Casas and Aurelia Guy and Arthur Mensch and Michela Paganini and Jordan Hoffmann and Bogdan Damoc and Blake Hechtman and Trevor Cai and Sebastian Borgeaud and George van den Driessche and Eliza Rutherford and Tom Hennigan and Matthew Johnson and Katie Millican and Albin Cassirer and Chris Jones and Elena Buchatskaya and David Budden and Laurent Sifre and Simon Osindero and Oriol Vinyals and Jack Rae and Erich Elsen and Koray Kavukcuoglu and Karen Simonyan",
    title       =   "Unified Scaling Laws for Routed Language Models",
    journal     =   "arXiv Preprint",
    volume      =   "2202.01169",
    year        =   "2022",
    pages       =   "1--31",
    note        =   "\url{https://arxiv.org/abs/2202.01169}",
}

@article{LaCroix-Mohseni-2022,
    author      =   "Travis LaCroix and Aydin Mohseni",
    title       =   "The tragedy of the AI commons",
    journal     =   "Synthese",
    volume      =   "200",
    number      =   "289",
    year        =   "2022", 
}

@book{Cohen-1972,
    author      =   "Stanley Cohen",
    title       =   "Folk Devils and Moral Panics: The Creation of the Mods and Rockers",
    publisher   =   "Granada Publishing",
    address     =   "London and Manchester, UK",
    year        =   "1972", 
}

@book{Costanza-Chock-2020,
    author      =   "Sasha Costanza-Chock",
    title       =   "Design Justice: Community-Led Practices to Build the Worlds We Need",
    publisher   =   "The MIT Press",
    address     =   "Cambridge, MA",
    year        =   "2020", 
}

@article{ET-Now-2025,
    author      =   "ET Now",
    title       =   "{EXCLUSIVE} | `Lot Of Enthusiasm Over {ChatGPT} From India': {OpenAI CEO Sam Altman} | {ET Now}",
    journal     =   "YouTube",
    volume      =   "Jun.",
    number      =   "09",
    year        =   "2023",
    notes       =   "\url{}",
}

@misc{Gallup2026,
  author =       "Gallup",
  year =         "2026",
  month =        May,
  title =        "Americans Oppose AI Data Centers in Their Area",
  date = "May 13, ",
  url =          "https://news.gallup.com/poll/709772/americans-oppose-data-centers-area.aspx",
}

@article{Gebru-et-al-2023,
    author      =   "Timnit Gebru and Emily M. Bender and Angelina McMillan-Major",
    title       =   "Statement from the listed authors of {Stochastic Parrots} on the `{AI} pause' letter",
    journal     =   "Distributed AI Research (DAIR) Institute",
    year        =   "2023", 
    volume      =   "Mar.",
    number      =   "31",
    note        =   "\url{https://dair-institute.org/blog/letter-statement-March2023/}",
    pages       =   "n.p.",
}

@book{Hall-et-al-1978,
    author      =   "Stuart Hall and Chas Critcher and Tony Jefferson and John Clarke and Brian Roberts",
    title       =   "Policing the Crisis: Mugging, the State, and Law and Order",
    publisher   =   "Bloomsbury",
    year        =   "1978", 
}

@book{Hasinoff-2015,
    author      =   "Amy Adele Asinoff",
    title       =   "Sexting Panic: Rethinking Criminalization, Privacy, and Consent",
    publisher   =   "University of Illinois Press",
    address     =   "Champaign, IL",
    year        =   "2015",
}

@article{Hoffmann-et-al-2022,
    author      =   "Jordan Hoffmann and Sebastian Borgeaud and Arthur Mensch and Elena Buchatskaya and Trevor Cai and Eliza Rutherford and Diego de Las Casas and Lisa Anne Hendricks and Johannes Welbl and Aidan Clark and Tom Hennigan and Eric Noland and Katie Millican and George van den Driessche and Bogdan Damoc and Aurelia Guy and Simon Osindero and Karen Simonyan and Erich Elsen, Jack W. Rae and Oriol Vinyals and Laurent Sifre",
    title       =   "Training Compute-Optimal Large Language Models",
    journal     =   "arXiv Preprint",
    volume      =   "2203.15556",
    year        =   "2022", 
    pages       =   "1--36",
    note        =   "\url{https://arxiv.org/abs/2203.15556}",
}

@article{Holden-Tanenbaum-2023,
    author      =   "LaTasha R. Holden and Gabriel J. Tanenbaum",
    title       =   "Modern Assessments of Intelligence Must Be Fair and Equitable",
    journal     =   "J Intell.",
    volume      =   "11",
    number      =   "6",
    year        =   "2023", 
    pages       =   "126",
}

@book{Johnson-2004,
    author      =   "David K. Johnson",
    title       =   "The Lavender Scare",
    publisher   =   "The University of Chicago Press",
    address     =   "Chicago",
    year        =   "2004", 
}

@article{Kaplan-et-al-2020,
    author      =   "Jared Kaplan and Sam McCandlish and Tom Henighan and Tom B. Brown and Benjamin Chess and Rewon Child and Scott Gray and Alec Radford and Jeffrey Wu and Dario Amodei",
    title       =   "Scaling Laws for Neural Language Models",
    journal     =   "arXiv Preprint",
    volume      =   "2001.08361",
    year        =   "2020",
    pages       =   "1--30",
    note        =   "\url{https://arxiv.org/abs/2001.08361}",
}

@book{LaCroix-2025,
    author      =   "Travis LaCroix",
    title       =   "Artificial Intelligence and the Value Alignment Problem: A Philosophical Introduction",
    publisher   =   "Broadview Press",
    address     =   "Peterborough, ON",
    year        =   "2025", 
}

@inproceedings{LaCroix-2026,
    author      =   "Travis LaCroix",
    title       =   "Relative Principals, Pluralistic Alignment, and the Structural Value Alignment Problem",
    booktitle   =   "{FAccT} '26: The 2026 {ACM} Conference on Fairness, Accountability, and Transparency",
    year        =   "2026", 
    pages       =   "622--639",
    notes       =   "\url{https://doi.org/10.1145/3805689.3812420}",
}

@inproceedings{LaCroix-et-al-2026,
    author      =   "Travis LaCroix and Fintan Mallory and Sasha Luccioni",
    title       =   "Strategic Polysemy in AI Discourse: A Philosophical Analysis of Language, Hype, and Power",
    booktitle   =   "{FAccT} '26: The 2026 {ACM} Conference on Fairness, Accountability, and Transparency",
    year        =   "2026", 
    pages       =   "497--517",
    notes       =   "\url{https://doi.org/10.1145/3805689.3812399}",
}

@article{LaCroix-2026-Autism,
    author      =   "Travis LaCroix",
    title       =   "Autism and the Pseudoscience of Mind",
    journal     =   "Psychological Inquiry",
    volume      =   "36",
    number      =   "4",
    year        =   "2025", 
    pages       =   "229--261",
}

@techreport{Legg-Hutter-2007,
    author      =   "Shane Legg and Marcus Hutter",
    title       =   "A Collection of Definitions of Intelligence",
    institution =   "IDSIA-07-07",
    year        =   "2007",
}

@article{Lord-2021,
    author      =   "Phil Lord",
    title       =   "Moral panic and the war on drugs",
    journal     =   "University of New Hampshire Law Review",
    volume      =   "20",
    year        =   "2021", 
    pages       =   "407",
}

@article{Marche-2022,
    author      =   "Stephen Marche",
    title       =   "The College Essay Is Dead",
    journal     =   "The Atlantic",
    volume      =   "Dec.",
    number      =   "06",
    year        =   "2022",
    note        =   "\url{https://www.theatlantic.com/technology/archive/2022/12/chatgpt-ai-writing-college-student-essays/672371/}",
}

@article{McInerney-2024,
    author      =   "Kerry McInerney",
    title       =   "Yellow Techno-Peril: The `Clash of Civilizations' and anti-Chinese racial rhetoric in the US–China AI arms race",
    journal     =   "Big Data \& Society",
    volume      =   "11",
    number      =   "2",
    year        =   "2024", 
    pages       =   "1--13",
}

@article{Morder-2026,
    author      =   "Robi Morder",
    title       =   " Why the student protests in France and the government’s response are unprecedented and unpredictable",
    journal     =   "The Conversation",
    volume      =   "Oct.",
    number      =   "07",
    year        =   "2026", 
    pages       =   "n.p.",
    note        =   "\url{https://theconversation.com/why-the-student-protests-in-france-and-the-governments-response-are-unprecedented-and-unpredictable-293863}",
}

@article{Novozhilova-et-al-2026,
    author      =   "Ekaterina Novozhilova and Chi Vu and James Katz",
    title       =   "From moral panic to normalization: Comparing users and non-users of AI companionship apps",
    journal     =   "{AI} \& Society",
    volume      =   "41",
    number      =   "4",
    year        =   "2026", 
    pages       =   "4057--4075",
}

@article{OBrien-2026,
    author      =   "Hettie O'Brien",
    title       =   "What's behind the global backlash against datacentres?",
    journal     =   "The Guardian",
    volume      =   "Sep.",
    number      =   "02",
    year        =   "2026", 
    note        =   "\url{https://www.theguardian.com/world/2026/sep/02/wednesday-briefing-whats-behind-the-global-backlash-against-datacentres}",
}

@article{Ohio-State-2025,
    author      =   "{The Ohio State University}",
    title       =   "Ohio State launches bold {AI} Fluency initiative to redefine learning and innovation",
    journal     =   "Ohio State News",
    volume      =   "Jun.",
    number      =   "04",
    year        =   "2025", 
    note        =   "\url{https://news.osu.edu/ohio-state-launches-bold-ai-fluency-initiative-to-redefine-learning-and-innovation/}",
}

@article{Pepin-Neff-2023,
    author      =   "Chris Pepin-Neff",
    title       =   "Anti-Trans Moral Panics Endanger All Young People",
    journal     =   "Scientific American",
    volume      =   "May",
    number      =   "19",
    year        =   "2023",
    note        =   "\url{https://www.scientificamerican.com/article/anti-trans-moral-panics-endanger-all-young-people/}",
}

@article{Rose-2023,
    author      =   "Steve Rose",
    title       =   "Five ways {AI} could improve the world: ‘We can cure all diseases, stabilise our climate, halt poverty’",
    journal     =   "The Guardian",
    volume      =   "Jul.",
    number      =   "06",
    year        =   "2023",
    pages       =   "n.p.",
    note        =   "\url{https://www.theguardian.com/technology/2023/jul/06/ai-artificial-intelligence-world-diseases-climate-scenarios-experts}",
}

@misc{Register2026,
  author =       "Vigliarolo, Brandon",
  year =         "2026",
  month =        "July",
  title =        "Amsterdam activists throw acid at Microsoft datacenter project",
  date = "July 16, ",
  journal= "The Register",
  url =          "https://www.theregister.com/ai-and-ml/2026/07/16/amsterdam-activists-throw-acid-at-microsoft-datacenter-project/5273943",
}

@article{Sutton-2019,
    author      =   "Rich Sutton",
    title       =   "The Bitter Lesson",
    journal     =   "Incomplete Ideas",
    year        =   "2019",
    note        =   "\url{http://www.incompleteideas.net/IncIdeas/BitterLesson.html}",
}

@article{Synergy-RG-2026,
    author      =   "{Synergy Research Group}",
    title       =   "Hyperscale Operators to Account for 67\% of all Data Center Capacity by 2031",
    journal     =   "SRG Research",
    volume      =   "Apr.",
    number      =   "07",
    year        =   "2026",
    note        =   "\url{https://www.srgresearch.com/articles/hyperscale-operators-to-account-for-67-of-all-data-center-capacity-by-2031}",
    pages       =   "n.p.",
}

@article{Taylor-LaCroix-2026,
    author      =   "Tristan B. Taylor and Travis LaCroix",
    title       =   "Purpose before policy: academic integrity, generative AI, and rhetorical stance",
    journal     =   "Higher Education",
    year        =   "2026", 
    pages       =   "1--20",
}

@article{van-den-Berg-2026,
    author      =   "Amy van den Berg ",
    title       =   "Mississauga city council votes to block data centre projects for 1 year",
    journal     =   "CBC",
    year        =   "2026",
    volume      =   "Sep.",
    number      =   "16",
    note        =   "\url{https://www.cbc.ca/news/canada/toronto/mississauga-data-centres-coucil-vote-passes-9.7346133}", 
}

@article{Whittaker-2021,
    author      =   "Meredith Whittaker",
    title       =   "The Steep Cost of Capture",
    journal     =   "Interactions",
    volume      =   "28",
    number      =   "6",
    year        =   "2021", 
    pages       =   "50--55",
}

@article{Wippell-et-al-2026,
    author      =   "Jack G. R. Wippell and Hollie Nyseth Nzitatira and Shelby Frye and Bri Londer{\'e}e and Ryleigh Adkins and Shreya Yogesh",
    title       =   "Creating an Anti-Transgender Moral Panic",
    journal     =   "Contexts",
    volume      =   "25",
    number      =   "2",
    year        =   "2026",
    pages       =   "28--35",
}

@article{Wood-et-al-2026,
    author      =   "Nathan Wood and Scott Robbins and Eduardo Zegarra Berodt and Anton Graf von Westerholt and Michelle Behrndt and Hauke Budig and Daniel Kloock-Schreiber",
    title       =   "Stop Saying `{AI}'",
    journal     =   "arXiv Preprint",
    volume      =   "2602.17729",
    year        =   "2026",
    pages       =   "1--49",
    note        =   "\url{https://arxiv.org/abs/2602.17729}",
}

@misc{NYState2026,
  author =       "{State of New York}",
  year =         "2026",
  month =        "July",
  title =        "First Statewide Moratorium on New Hyperscale Data Centers Launched by Governor Kathy Hochul",
  date = "July 14, ",
  url =  "https://www.governor.ny.gov/news/first-statewide-moratorium-new-hyperscale-data-centers-launched-governor-kathy-hochul",
}

@misc{Bloomberg2026,
  author =       "Anunchit Nguyen",
  year =         "2026",
  month =        "September",
  title =        "Thailand Puts 49 Data Centers on Hold Over Resource Strain",
  date = "September 4, ",
  journal = "Bloomberg",
  url =  "http://bloomberg.com/news/articles/2026-09-04/thailand-puts-49-data-centers-on-hold-over-resource-strain",
}

@misc{Bloomberg2026_2,
  author =       "Alicia Clanton",
  year =         "2026",
  month =        "October",
  title =        "San Francisco Passes Data Center Moratorium in Hometown of AI",
  date = "October 10,",
  journal = "Bloomberg",
  url =  "https://www.bloomberg.com/news/articles/2026-10-07/san-francisco-passes-data-center-moratorium-in-hometown-of-ai",
}

@misc{techpolicypress2026,
  author =       "Varsha Bansal",
  year =         "2026",
  month =        "June",
  title =        "Tech Workers Are Fighting Against Silicon Valley’s AI Push",
  date = "June 17, ",
  journal = "Tech Policy Press",
  url =  "https://www.techpolicy.press/tech-workers-are-fighting-against-silicon-valleys-ai-push/",
}

@misc{gulfnews2026,
  author =       "Jay Hilotin",
  year =         "2026",
  month =        "September",
  title =        "Elon Musk: AI could usher in an ‘age of abundance’ for everyone — what critics are saying",
  date = "September 30, ",
  journal = "Gulf News",
  url =  "https://gulfnews.com/technology/elon-musk-ai-could-usher-in-an-age-of-abundance-for-everyone-what-critics-are-saying-1.500692945",
}

@misc{altman2025,
  author =       "Sam Altman",
  year =         "2025",
  month =        "September",
  title =        "Abundant Intelligence",
  date = "September 23, 2025",
  url =  "https://blog.samaltman.com/abundant-intelligence",
}

@misc{altman2024,
  author =       "Sam Altman",
  year =         "2024",
  title =        "The Intelligence Age",
  month = "September 23,",
  url =  "https://ia.samaltman.com/",
}

@misc{amodei2024,
  author =       "Dario Amodei",
  year =         "2024",
  month =        "October",
  title =        "TMachines of Loving Grace: How AI Could Transform the World for the Better",
  date = "October 2024",
  url = "https://darioamodei.com/essay/machines-of-loving-grace",
}

@article{beebe2007empirical,
  title={An empirical study of US copyright fair use opinions, 1978-2005},
  author={Beebe, Barton},
  journal={U. Pa. L. Rev.},
  volume={156},
  pages={549},
  year={2007},
  publisher={HeinOnline}
}

@inproceedings{jiang2023,
author = {Jiang, Harry H. and Brown, Lauren and Cheng, Jessica and Khan, Mehtab and Gupta, Abhishek and Workman, Deja and Hanna, Alex and Flowers, Johnathan and Gebru, Timnit},
title = {AI Art and its Impact on Artists},
year = {2023},
isbn = {9798400702310},
publisher = {Association for Computing Machinery},
address = {New York, NY, USA},
url = {https://doi.org/10.1145/3600211.3604681},
doi = {10.1145/3600211.3604681},
booktitle = {Proceedings of the 2023 AAAI/ACM Conference on AI, Ethics, and Society},
pages = {363–374},
numpages = {12},
location = {Montr{\'e}al, QC, Canada},
series = {AIES '23}
}

@inproceedings{kawakami2024,
author = {Kawakami, Reishiro and Venkatagiri, Sukrit},
title = {The Impact of Generative AI on Artists},
year = {2024},
isbn = {9798400704857},
publisher = {Association for Computing Machinery},
address = {New York, NY, USA},
url = {https://doi.org/10.1145/3635636.3664263},
doi = {10.1145/3635636.3664263},
booktitle = {Proceedings of the 16th Conference on Creativity \& Cognition},
pages = {79–82},
numpages = {4},
location = {Chicago, IL, USA},
series = {C\&C '24}
}

@article{Klein-2026,
    author      =   "Ezra Klein",
    title       =   "Why the {A.I.} Job Apocalypse (Probably) Won’t Happen",
    journal     =   "The New York Times",
    volume      =   "May.",
    number      =   "23",
    year        =   "2026", 
    pages       =   "n.p.",
    note        =   "\url{https://www.nytimes.com/2026/05/03/opinion/ai-jobs-unemployment-silicon-valley.html}",
}

@misc{amodei2026,
  author =       "Dario Amodei",
  year =         "2026",
  month =        "September",
  title =        "We Must Pace the Frontier",
  date = "September 2026",
  url =  "https://darioamodei.com/post/we-must-pace-the-frontier",
}

@misc{cole2023,
  author =       "Samantha Cole",
  year =         "2023",
  month =        "February",
  title =        "Netflix Made an Anime Using AI Due to a ’Labor Shortage,’and Fans Are Pissed",
  date = "February 1 2023",
  url =  "https://www.vice.com/en/article/netflix-anime-dog-and-the-boy-ai-generated-art/",
}

@misc{lang2026,
  author =       "Jamie Lang",
  year =         "2026",
  month =        "October",
  title =        "China’s First AI Feature Flops, Months After Scrappy ‘Niu Lai’ Became A Box Office Hit",
  lastaccessed = "October 26 2026",
  url =  "https://www.cartoonbrew.com/box-office-report/chinese-ai-film-sanxingdui-future-past-box-office-flop-268426.html",
}

@misc{hayes2026,
  author =       "Dade Hayes",
  year =         "2026",
  month =        May,
  title =        "Disney’s Recent Layoffs May Not Be One-Offs As Company Touts “Culture Of Efficiency”",
  date = "May 6",
  url =  "https://deadline.com/2026/05/disney-layoffs-workers-ai-culture-of-efficiency-1236882815/",
}

@misc{vallance2022,
  author =       "Chris Vallance",
  year =         "2022",
  month =        "October",
  title =        "``Art is dead Dude`` - the rise of the AI artists stirs debate",
  date = "13 September ",
  journal = "BBC",
  url =  "https://www.bbc.com/news/technology-62788725",
}

@misc{landrymore2024,
  author =       "Frank Landymore",
  year =         "2024",
  month =        "June",
  title =        "OpenAI Exec Says AI Will Kill Creative Jobs That “Shouldn’t Have Been There in the First Place”",
  date = "Jun 25,",
  journal = "Futurism",
  url =  "https://futurism.com/the-byte/openai-exec-ai-creative-jobs-shouldnt-exist",
}

@misc{shankland2023,
  author =       "Stephen Shankland",
  year =         "2023",
  month =        "March",
  title =        "Adobe Says Its New Generative AI Will Help Creative Pros, Not Hurt Them",
  date = "March 23,",
  journal = "CNET",
  url =  "https://www.cnet.com/tech/computing/adobe-says-its-new-generative-ai-will-help-creative-pros-not-hurt-them/",
}

@misc{merchant2025,
  author =       "Brian Merchant",
  year =         "2025",
  month =        "September",
  title =        "Artists are losing work, wages, and hope as bosses and clients embrace AI",
  date = "September 16,",
  journal = "Blood in the Machine",
  url =  "https://www.bloodinthemachine.com/p/artists-are-losing-work-wages-and",
}

@article{zhou2024generative,
  title={Generative artificial intelligence, human creativity, and art},
  author={Zhou, Eric and Lee, Dokyun},
  journal={PNAS nexus},
  volume={3},
  number={3},
  pages={pgae052},
  year={2024},
  publisher={Oxford University Press, US}
}

@inproceedings{jiang2026professional,
  title={How Professional Visual Artists are Negotiating Generative AI in the Workplace},
  author={Jiang, Harry H and Taylor, Jordan and Agnew, William},
  booktitle={Proceedings of the Extended Abstracts of the 2026 CHI Conference on Human Factors in Computing Systems},
  pages={1--6},
  year={2026}
}

@article{razak2026distinguishing,
  title={Distinguishing Human From Machine: High School Students’ Identification and Perception of AI Art},
  author={Razak, Nuha},
  journal={Critical Debates in Humanities, Science and Global Justice},
  year={2026}
}

@article{sanlit2025,
author = {Deniz Esin Tekcan Sanli and Ahmet Necati Sanli and Duzgun Yildirim and Ilkay Dogan},
title ={Can ChatGPT detect breast cancer on mammography?},
journal = {Journal of Medical Screening},
volume = {32},
number = {3},
pages = {172-175},
year = {2025},
doi = {10.1177/09691413251334587},
    note ={PMID: 40259576},
URL = { https://doi.org/10.1177/09691413251334587
},
}

@article{arravalli2025,
  title={Detection of breast cancer using machine learning and explainable artificial intelligence},
  author={Arravalli, Tharunya and Chadaga, Krishnaraj and Muralikrishna, H and Sampathila, Niranjana and Cenitta, D and Chadaga, Rajagopala and Swathi, KS},
  journal={Scientific Reports},
  volume={15},
  number={1},
  pages={26931},
  year={2025},
  publisher={Nature Publishing Group UK, London}
}

@article{van2025human,
  title={Human perception of art in the age of artificial intelligence},
  author={Van Hees, Jules and Grootswagers, Tijl and Quek, Genevieve L and Varlet, Manuel},
  journal={Frontiers in psychology},
  volume={15},
  pages={1497469},
  year={2025},
  publisher={Frontiers}
}

@misc{fortune2026,
  author =       "Tristan Bove",
  year =         "2026",
  month =        "February",
  title =        "Big Tech execs playing ‘Russian roulette’ in the AI arms race could risk human extinction, warns top researcher",
  magazine = "Fortune",
  date = "February 16,",
  url =  "https://fortune.com/2026/02/18/big-tech-russian-roulette-ai-race-humanity-extinction/",
}

@misc{canada2026,
  author =       {Government of Canada},
  year =         "2026",
  month =        "September",
  title =        "Government of Canada launches National AI Literacy Initiative",
  magazine = "Innovation, Science and Economic Development Canada",
  date = "September 9,",
  url =  "https://www.canada.ca/en/innovation-science-economic-development/news/2026/09/government-of-canada-launches-national-ai-literacy-initiative.html",
}

@misc{cress226,
  author =       {Laura Cress},
  year =         "2026",
  month =        "January",
  title =        "Government offers UK adults free AI training for work",
  magazine = "BBC",
  date = "28 January,",
  url =  "https://www.bbc.com/news/articles/cp37prvp072o",
}

@misc{rogers2024,
      title={Position: Key Claims in LLM Research Have a Long Tail of Footnotes}, 
      author={Anna Rogers and Alexandra Sasha Luccioni},
      year={2024},
      eprint={2308.07120},
      archivePrefix={arXiv},
      primaryClass={cs.CL},
      url={https://arxiv.org/abs/2308.07120}, 
}

@article{Trump-2026,
    author      =   "Donald J. Trump",
    title       =   "INAUGURATING THE ERA OF SUPER INTELLIGENCE",
    journal     =   "The White House, Executive Order I4434",
    volume      =   "Sep.",  
    number      =   "29",
    year        =   "2026",
    pages       =   "n.p.",
    note        =   "\url{https://www.whitehouse.gov/presidential-actions/2026/09/inaugurating-the-era-of-super-intelligence/}",
}

@book{Williams-1985,
    author      =   "Bernard Williams",
    title       =   "Ethics and the Limits of Philosophy",
    publisher   =   "Harvard University Press",
    address     =   "Cambridge, MA",
    year        =   "1985", 
}

@article{alanezi2026,
  title={The influence of AI literacy on fear of negative evaluation among students and researchers: a multisite cross-sectional study},
  author={Alanezi, Fahad Zeed and Alenazi, Latifah H and Alkathiri, Alwah Mohammed and Innab, Adnan},
  journal={BMC Medical Education},
  volume        =   "",
  number        =   "",
  year={2026},
  publisher={Springer}
}

@article{lee2026mediating,
  title={Mediating effect of technostress on the relationship between artificial intelligence literacy and attitude toward digital technology among health professions students: a structural equation modeling approach},
  author={Lee, Jin Young and Min, Yul Ha and Yim, Jun and Park, Kwi Hwa and Yune, So Jung},
  journal={Journal of Yeungnam Medical Science},
  volume={43},
  year={2026},
  publisher={Journal of Yeungnam Medical Science}
}

@article{satmaz2026ai,
  title={AI literacy is directly associated with positive attitudes and indirectly associated with negative attitudes through AI anxiety: a cross-sectional study of preservice teachers},
  author={Satmaz, {\.I}smail and Ad{\i}yaman, Mehmet},
  journal={Frontiers in Psychology},
  volume={17},
  pages={1946898},
  year={2026},
  publisher={Frontiers Media SA}
}

@misc{jacobin2026,
  author =       "Noah Gordon",
  year =         "2026",
  month =        "October",
  title =        "Could AI Fix Climate Change? Probably Not Under Capitalism.",
  magazine = "Jacobin",
  date = "October 07, ",
  url =  "https://jacobin.com/2026/10/ai-huang-climate-change-capitalism",
}

@misc{collectiveintelligence2024,
  author =       "Noah Gordon",
  year =         "2024",
  month =        "March",
  title =        "A Roadmap to
Democratic AI",
  date = "March",
  url =  "https://static1.squarespace.com/static/631d02b2dfa9482a32db47ec/t/65f9a1296f1a357e918f7a58/1722968408230/CIP_+A+Roadmap+to+Democratic+AI.pdf",
}

@article{avery2026,
  title={Introducing the" Hard" Problem of Artificial Legal Intelligence},
  author={Avery, Joseph J},
  journal={NYUL Rev. Online},
  volume={101},
  pages={116},
  year={2026},
  publisher={HeinOnline}
}

@article{baban2025,
  title={Artificial Intelligence and the Future of Legal Accountability},
  author={Baban More, Dr Rohit Jadhav and Madan, Mr Sunil and Munjankar, Mr Vijaydeep},
  year={2025}
}

@article{veltman2025,
  title={Anthropic settles with authors in first-of-its-kind AI copyright infringement lawsuit},
  author={Chloe Veltman},
  year={2025},
  date="September 5, 2025", url="https://www.npr.org/2025/09/05/nx-s1-5529404/anthropic-settlement-authors-copyright-ai"
}

@article{cecco2024,
  title={Air Canada ordered to pay customer who was misled by airline’s chatbot},
  author={Leyland Cecco},
  year={2024},
  journal = "The Guardian",
  date="February 16 2024", url="https://www.theguardian.com/world/2024/feb/16/air-canada-chatbot-lawsuit"
}

@article{libcongress2026,
  title={Germany: Court Rules Chatbot Operators Are Liable for AI Hallucinations},
  author={Laura Baitz},
  year={2026},
  journal = "Law Library of Congress",
  date="June 9, 2026", url="https://www.loc.gov/item/global-legal-monitor/2026-06-09/germany-court-rules-chatbot-operators-are-liable-for-ai-hallucinations/"
}

@article{colorado2026,
  title={Colorado SB26-189
Automated Decision-Making Technology},
  author={Colorado General Assembly},
  year={2026},
  journal = "2026 Regular Session",
  date="2026", url="https://leg.colorado.gov/bills/sb26-189"
}

@article{eo2023,
  title={Safe, Secure, and Trustworthy Development and Use of Artificial Intelligence},
  author={United States Government},
  year={2023},
  journal = "Executive Office of the President of the United States",
  date="October 30, 2023", url="https://www.federalregister.gov/documents/2023/11/01/2023-24283/safe-secure-and-trustworthy-development-and-use-of-artificial-intelligence"
}

@article{openai2026,
  title={Our approach to EU text provenance rules},
  author={OpenAI},
  year={2026},
  journal = "Open AI Blog",
  date="October 5, 2026", url="https://openai.com/index/eu-text-provenance/"
}

@article{anthropic2026,
  title={How Claude’s text watermark works},
  author={Anthropic},
  year={2026},
  journal = "Anthropic Blog",
  date="Aug 14, 2026", url="https://www.anthropic.com/news/claude-text-watermark"
}

@article{dorm2026,
title = {Large language models possess some ecological knowledge, but how much?},
journal = {Ecological Informatics},
volume = {95},
pages = {103699},
year = {2026},
issn = {1574-9541},
doi = {https://doi.org/10.1016/j.ecoinf.2026.103699},
url = {https://www.sciencedirect.com/science/article/pii/S1574954126001056},
author = {Filip Dorm and Joseph Millard and Drew Purves and Michael Harfoot and Oisin {Mac Aodha}},

}

@article{courty2023mlco2,
  title={mlco2/codecarbon: v3. 2.1},
  author={Courty, Benoit and Schmidt, Victor and Blanche, Luis and Feld, Boris and Lecourt, J{\'e}r{\'e}my and Saboni, Amine and LLORET, Patrick and L{\'e}val, Mathilde and Cruveiller, Alexis and Zhao, Franklin and others},
  journal={Zenodo},
  year={2023}
}

@inproceedings{mitchell2019,
  title={Model cards for model reporting},
  author={Mitchell, Margaret and Wu, Simone and Zaldivar, Andrew and Barnes, Parker and Vasserman, Lucy and Hutchinson, Ben and Spitzer, Elena and Raji, Inioluwa Deborah and Gebru, Timnit},
  booktitle={Proceedings of the conference on fairness, accountability, and transparency},
  pages={220--229},
  year={2019}
}

@article{gebru2021,
  title={Datasheets for datasets},
  author={Gebru, Timnit and Morgenstern, Jamie and Vecchione, Briana and Vaughan, Jennifer Wortman and Wallach, Hanna and Iii, Hal Daum{\'e} and Crawford, Kate},
  journal={Communications of the ACM},
  volume={64},
  number={12},
  pages={86--92},
  year={2021},
  publisher={ACM, New York, NY, USA}
}

@book{cant2024feeding,
  title={Feeding the machine: The hidden human labor powering AI},
  author={Cant, Callum and Muldoon, James and Graham, Mark},
  year={2024},
  publisher={Bloomsbury Publishing USA}
}

@misc{miceli2022,
      title={Documenting Data Production Processes: A Participatory Approach for Data Work}, 
      author={Milagros Miceli and Tianling Yang and Adriana Alvarado Garcia and Julian Posada and Sonja Mei Wang and Marc Pohl and Alex Hanna},
      year={2022},
      eprint={2207.04958},
      archivePrefix={arXiv},
      primaryClass={cs.HC},
      url={https://arxiv.org/abs/2207.04958}, 
}

@misc{luccioni2025,
      title={Misinformation by Omission: The Need for More Environmental Transparency in AI}, 
      author={Sasha Luccioni and Boris Gamazaychikov and Theo Alves da Costa and Emma Strubell},
      year={2025},
      eprint={2506.15572},
      archivePrefix={arXiv},
      primaryClass={cs.CY},
      url={https://arxiv.org/abs/2506.15572}, 
}

@article{dathathri2024scalable,
  title={Scalable watermarking for identifying large language model outputs},
  author={Dathathri, Sumanth and See, Abigail and Ghaisas, Sumedh and Huang, Po-Sen and McAdam, Rob and Welbl, Johannes and Bachani, Vandana and Kaskasoli, Alex and Stanforth, Robert and Matejovicova, Tatiana and others},
  journal={Nature},
  volume={634},
  number={8035},
  pages={818--823},
  year={2024},
  publisher={Nature Publishing Group UK, London}
}

@techreport{li2026textgrain,
  title={textGrain: Entropy-calibrated watermarking for language model text},
  author={Li, Xiang and Wen, Garrett and Chen, Xiaohong and Long, Qi and Jain, Arzav and Joly, Florent and Lam, Mike and Song, Qingquan and Su, Weijie},
  year={2026},
  institution={Technical report, OpenAI},
  url="https://cdn.openai.com/pdf/e9508624-d767-41b6-a26d-e34ca798ada6/textgrain-entropy-calibrated-watermarking-for-language-model-text.pdf"
}

@article{moats2025,
title = {Bringing AI participation down to scale},
journal = {Patterns},
volume = {6},
number = {5},
pages = {101241},
year = {2025},
issn = {2666-3899},
doi = {https://doi.org/10.1016/j.patter.2025.101241},
url = {https://www.sciencedirect.com/science/article/pii/S2666389925000893},
author = {David Moats and Chandrima Ganguly},
}

@misc{ovadya2025,
      title={Democratic AI is Possible. The Democracy Levels Framework Shows How It Might Work}, 
      author={Aviv Ovadya and Kyle Redman and Luke Thorburn and Quan Ze Chen and Oliver Smith and Flynn Devine and Andrew Konya and Smitha Milli and Manon Revel and K. J. Kevin Feng and Amy X. Zhang and Bilva Chandra and Michiel A. Bakker and Atoosa Kasirzadeh},
      year={2025},
      eprint={2411.09222},
      archivePrefix={arXiv},
      primaryClass={cs.CY},
      url={https://arxiv.org/abs/2411.09222}, 
}

@article{tseng2023lightweight,
  title={Lightweight, pre-trained transformers for remote sensing timeseries},
  author={Tseng, Gabriel and Cartuyvels, Ruben and Zvonkov, Ivan and Purohit, Mirali and Rolnick, David and Kerner, Hannah},
  journal={arXiv preprint arXiv:2304.14065},
  year={2023}
}

@article{aryafar2022,
  title={AI could help realize the dream of the four-day work week},
  author={Kamelia Aryafar},
  journal={Google Cloud Blog},
  url= "https://cloud.google.com/blog/products/ai-machine-learning/it-prediction-ai-could-help-realize-the-dream-of-the-four-day-work-week",
  year={2022}
}

@incollection{jevons1865coal,
  title={The Coal Question: An Inquiry Concerning the Progress of the Nation, and the Probable Exhaustion of Our Coal-Mines},
  author={Jevons, Stanley W},
  year={1865}
}

@article{patterson2021carbon,
  title={Carbon emissions and large neural network training},
  author={Patterson, David and Gonzalez, Joseph and Le, Quoc and Liang, Chen and Munguia, Lluis-Miquel and Rothchild, Daniel and So, David and Texier, Maud and Dean, Jeff},
  journal={arXiv preprint arXiv:2104.10350},
  year={2021}
}

@article{lacroix2025metaethical,
  title={Metaethical perspectives on ‘benchmarking’AI ethics},
  author={LaCroix, Travis and Luccioni, Alexandra Sasha},
  journal={AI and Ethics},
  volume={5},
  number={4},
  pages={4029--4047},
  year={2025},
  publisher={Springer}
}

@inproceedings{luccioni2024power,
  title={Power hungry processing: Watts driving the cost of AI deployment?},
  author={Luccioni, Sasha and Jernite, Yacine and Strubell, Emma},
  booktitle={Proceedings of the 2024 ACM conference on fairness, accountability, and transparency},
  pages={85--99},
  year={2024}
}

@article{pineau2021improving,
  title={Improving reproducibility in machine learning research (a report from the neurips 2019 reproducibility program)},
  author={Pineau, Joelle and Vincent-Lamarre, Philippe and Sinha, Koustuv and Larivi{\`e}re, Vincent and Beygelzimer, Alina and d'Alch{\'e}-Buc, Florence and Fox, Emily and Larochelle, Hugo},
  journal={Journal of machine learning research},
  volume={22},
  number={164},
  pages={1--20},
  year={2021}
}

@article{choudhary2022recent,
  title={Recent advances and applications of deep learning methods in materials science},
  author={Choudhary, Kamal and DeCost, Brian and Chen, Chi and Jain, Anubhav and Tavazza, Francesca and Cohn, Ryan and Park, Cheol Woo and Choudhary, Alok and Agrawal, Ankit and Billinge, Simon JL and others},
  journal={npj Computational Materials},
  volume={8},
  number={1},
  pages={59},
  year={2022},
  publisher={Nature Publishing Group UK London}
}

@article{rolnick2022tackling,
  title={Tackling climate change with machine learning},
  author={Rolnick, David and Donti, Priya L and Kaack, Lynn H and Kochanski, Kelly and Lacoste, Alexandre and Sankaran, Kris and Ross, Andrew Slavin and Milojevic-Dupont, Nikola and Jaques, Natasha and Waldman-Brown, Anna and others},
  journal={ACM computing surveys (CSUR)},
  volume={55},
  number={2},
  pages={1--96},
  year={2022},
  publisher={ACM New York, NY}
}

@misc{nyt2026_2,
  author =       "Ezra Klein",
  year =         "2026",
  month =        "October",
  title =        "Jensen Huang Thinks A.I. Alarmism Has Gone Too Far",
  magazine = "New York Times",
  date = "September 23,",
  url =  "https://www.nytimes.com/2026/09/23/opinion/ezra-klein-podcast-jensen-huang.html",
}

@misc{michael2026,
  author =       "Amelia Michael",
  year =         "2026",
  month =        "October",
  title =        "How did xAI build its Colossus data centers so fast?",
  date = "January 28,",
  url =  "https://ameliakmichael.substack.com/p/how-did-xai-build-its-colossus-data",
}

@misc{thomas2026,
  author =       "Michael Thomas",
  year =         "2026",
  month =        "October",
  title =        "Bypassing the Grid: How Data Center Developers Are Building Their Own Power Plants",
  date = "October 16,",
  url =  "https://cleanview.co/reports/behind-the-meter-data-centers",
}

@misc{politico2026,
  author =       "Ariel Wittenberg",
  year =         "2026",
  month =        "October",
  title =        "Anger Over Data Centers Is Fueling a New Political Movement",
  journal = "Politico",
  date = "October 2, ",
  url =  "https://www.politico.com/news/magazine/2026/08/02/the-new-luddites-01017824",
}

@misc{capitalb2026,
  author =       "Evan Simon",
  year =         "2026",
  month =        "February",
  title =        "Drone Footage Shows Musk’s AI Power Plant Flouting Clean Air Regulations in Black Community",
  journal = "Capital B News",
  date = "February 17,",
  url =  "https://capitalbnews.org/xai-musk-data-centers-clean-air-epa/",
}

@misc{taiwan2026act,
  title={Artificial Intelligence Basic Act},
  author={Taiwan National Science and Technology Council},
  url={https://law.nstc.gov.tw/EngLawContent.aspx?id=10099&lan=E},
  date={14 Jan 2026},
  year={2026}
}

@misc{uk2026act,
  title={UK Crime and Policing Act 2026},
  author={UK Public General Acts},
  url={https://www.legislation.gov.uk/ukpga/2026/20/contents},
  date={9 Sept 2026},
  year={2026}
}

@misc{kazakhstan2025act,
  title={Draft Law 'On Artificial Intelligence'},
  author={Kazakhstan Government},
  url={https://regulations.ai/regulations/RAI-KZ-NA-DAIPSXX-2025},
  date={19 Jan 2026},
  year={2026}
}

@misc{peru20235act,
  title={Peru - AI Promotion Law (31814)},
  author={Congress of Peru},
  url={https://www.gob.pe/institucion/congreso-de-la-republica/normas-legales/4565760-31814},
  date={5 July 2023},
  year={2023}
}

@misc{korea2026act,
  title={South Korea AI Trust Framework Act},
  author={Republic of Korea},
  url={https://elaw.klri.re.kr/eng_service/lawView.do?hseq=71019&lang=ENG},
  date={22 Jan 2026},
  year={2026}
}

@misc{us2020aiact,
  title={National Artificial Intelligence Initiative Act of 2020},
  author={116th Congress of the United States of America},
  url={https://www.congress.gov/bill/116th-congress/house-bill/6216},
  year={2020}
}

@misc{act2024eu,
  title={The {EU} artificial intelligence act},
  author={European Union},
  url={https://artificialintelligenceact.eu/},
  year={2024}
}

@article{winter2024governance,
  title={The Governance Misspecification Problem},
  author={Winter, Christoph and Bullock, Charlie},
  journal={Institute for Law \& AI Working Paper Series},
  volume={3},
  pages={1--29},
  year={2024}
}

@article{walz2019implementing,
  title={Implementing ethics into artificial intelligence: A contribution, from a legal perspective, to the development of an AI governance regime},
  author={Walz, Axel and Firth-Butterfield, Kay},
  journal={Duke L. \& Tech. Rev.},
  volume={18},
  pages={176},
  year={2019},
  publisher={HeinOnline}
}

@article{birmpili2018montreal,
  title={Montreal Protocol at 30: The governance structure, the evolution, and the Kigali Amendment},
  author={Birmpili, Tina},
  journal={Comptes Rendus Geoscience},
  volume={350},
  number={7},
  pages={425--431},
  year={2018},
  publisher={Elsevier}
}

@article{marchant2011growing,
  title={The growing gap between emerging technologies and the law},
  author={Marchant, Gary E},
  journal={The growing gap between emerging technologies and legal-ethical oversight: The pacing problem},
  pages={19--33},
  year={2011},
  publisher={Springer}
}

@book{collingridge1980social,
  title={The social control of technology},
  author={Collingridge, David},
  year={1980}
}

@article{birkstedt2023ai,
  title={AI governance: themes, knowledge gaps and future agendas},
  author={Birkstedt, Teemu and Minkkinen, Matti and Tandon, Anushree and M{\"a}ntym{\"a}ki, Matti},
  journal={Internet Research},
  volume={33},
  number={7},
  pages={133--167},
  year={2023},
  publisher={Emerald Publishing Limited}
}

@article{cath2018governing,
  title={Governing artificial intelligence: ethical, legal and technical opportunities and challenges},
  author={Cath, Corinne},
  journal={Philosophical transactions. Series A, Mathematical, physical, and engineering sciences},
  volume={376},
  number={2133},
  pages={20180080},
  year={2018}
}

@article{marchant2020governance,
  title={Governance of emerging technologies as a wicked problem},
  author={Marchant, Gary E},
  journal={Vanderbilt Law Revbiew},
  volume={73},
  pages={1861},
  year={2020},
  publisher={HeinOnline}
}

@misc{scientificamerican2026,
  author =       " Austyn Gaffney",
  year =         "2026",
  month =        "September",
  title =        "The data center backlash comes for SpaceXAI’s Colossus",
  journal = "Scientific American",
  date = "September 9,",
  url =  "https://www.scientificamerican.com/article/spacexais-colossus-data-center-is-fueling-a-massive-backlash",
}

@misc{ctv2026,
  author =       "Spencer Van Dyk",
  year =         "2026",
  month =        "September",
  title =        "Godfather of AI says it’s time to ‘wake up’ to the existential threat of AI",
  magazine = "CTV",
  date = "September 19,",
  url =  "https://www.ctvnews.ca/sci-tech/article/time-for-people-governments-to-wake-up-to-ai-existential-threat-bengio/",
}

@misc{fli2023,
  author =       "Future of Life Institute",
  year =         "2023",
  month =        "March",
  title =        "Pause Giant AI Experiments: An Open Letter",
  date = "22 March,",
  url =  "https://futureoflife.org/open-letter/pause-giant-ai-experiments/",
}

@misc{wsj2025,
  author =       "Josie Reich",
  year =         "2025",
  month =        "August",
  title =        "What Musk, Altman and Others Say About AI-Funded ‘Universal Basic Income’",
  date = "August 14, ",
  journal = "Wall Street Journal",
  url =  "https://www.wsj.com/tech/ai/universal-income-tech-executives-a16eb2d0",
}

@misc{bbc2023,
  author =       "Joe Tidy",
  year =         "2023",
  title =        "Worldcoin: Sam Altman launches eyeball scanning crypto coin",
  date = "24 July",
  journal = "BBC",
  url =  "https://www.bbc.com/news/technology-66128111",
}

@misc{bbc2026,
  author =       "Kali Hays",
  year =         "2026",
  title =        "Meta to cut one in 10 jobs after spending billions on AI",
  date = "24 April",
  journal = "BBC",
  url =  "https://www.bbc.com/news/articles/crm1y89vek8o",
}

@misc{hbr2026,
  author = "Aruna Ranganathan and Xingqi Maggie Ye",
  year =  "2026",
  title =  "AI Doesn’t Reduce Work—It Intensifies It",
  date = "February 9,",
  journal = "Harvard Business Review",
  url =  "https://hbr.org/2026/02/ai-doesnt-reduce-work-it-intensifies-it",
}

@misc{bloombergobcuts2023,
  author = "Subrat Patnaik and Ryan Vlastelica",
  year =  "2023",
  title =  "Big Tech’s Job Cuts Spur Rallies Even as an Economic Slowdown Looms",
  date = "January 25,",
  journal = "Bloomberg",
  url =  "https://www.bloomberg.com/news/articles/2023-01-25/big-tech-s-job-cuts-spur-rallies-even-as-economic-slowdown-looms",
}

@misc{cnbcjobcuts2026,
  author = "Liz Napolitano",
  year =  "2026",
  title =  "Markets AI-related layoffs a boost for stocks? Not necessarily",
  date = "May 17",
  journal = "CNBC",
  url =  "https://www.cnbc.com/2026/05/17/ai-related-layoffs-a-boost-for-stocks-not-necessarily.html",
}

@misc{challenger2026,
  author = "Challenger, Gray \& Christmas",
  year =  "2026",
  title =  "Challenger, Gray \& Christmas Job Cut Announcement Report",
  date = "August 6",
  url =  "https://www.challengergray.com/wp-content/uploads/2026/08/Challenger-Report-July-2026.pdf",
}

@article{acemoglu2026we,
  title={Why we must stop talking about artificial general intelligence-and instead build'pro-worker'AI},
  author={Acemoglu, Daron},
  journal={Nature},
  volume={656},
  number={8129},
  pages={795},
  year={2026}
}

@techreport{acemoglu2026building,
  title={Building pro-worker artificial intelligence},
  author={Acemoglu, Daron and Autor, David and Johnson, Simon},
  year={2026},
  institution={National Bureau of Economic Research}
}

@article{ding2026tracking,
  title={Tracking Artificial Intelligence Sentiment in the US Labor Market},
  author={Ding, Yuye and Ma, Mark Shuai and Wu, Jun and Yang, Yucheng},
  journal={Jun and Yang, Yucheng, Tracking Artificial Intelligence Sentiment in the US Labor Market (April 26, 2026)},
  year={2026}
}

@misc{sectionai2026,
  author =  "Section AI Blog",
  year =  "2026",
  title =  "The AI Proficiency Report",
  date = "June ",
  journal = "Section AI Blog",
  url =  "https://www.sectionai.com/ai/the-ai-proficiency-report",
}

@misc{andrew2023,
  author =  "Andrew Clement",
  year =  "2023",
  title =  "No AIDA is better than this AIDA",
  date = "November 8",
  journal = "A brief submitted to The Standing Committee on Industry and Technology (INDU) on Bill C-27,",
  url =  "https://www.ourcommons.ca/Content/Committee/441/INDU/Brief/BR12743452/br-external/ClementAndrew-e.pdf",
}

@misc{solomun2022,
  author =  "Sonja Solomun and Christelle Tessono",
  year =  "2022",
  title =  "How To Fix Canada’s Proposed Artificial Intelligence Act",
  date = "December",
  journal = "Tech Policy Press",
  url =  "https://www.techpolicy.press/how-to-fix-canadas-proposed-artificial-intelligence-act/",
}

@misc{satter2026,
  author =  "Raphael Satter and Deepa Seetharaman",
  year =  "2026",
  title =  "OpenAI's rogue agents probed Hugging Face for weaknesses two months before major hack",
  date = "September",
  journal = "Reuters",
  url =  "https://www.reuters.com/legal/litigation/openais-rogue-agents-probed-hugging-face-weaknesses-two-months-before-major-hack-2026-09-16/",
}

@misc{schroeder2026,
  author =  "Stan Schroeder",
  year =  "2026",
  title =  "OpenAI agent went rogue, escaped, and hacked Hugging Face",
  date = "July",
  journal = "Mashable",
  url =  "https://mashable.com/tech/hugging-face-openai-rogue-agent-hack-explained",
}

@misc{sekulich2026,
  author =  "Harry Sekulich and Lana Lam",
  year =  "2026",
  title =  "Rogue OpenAI agent 'infiltrated' Australian government website in world first",
  date = "September",
  journal = "BBC",
  url =  "https://www.bbc.com/news/articles/c6vgy0333dppo",
}

@misc{different_flops_2026,
      title={AI's Never Just One Thing: Different FLOPS for Different Folks},
      author={Jernite, Yacine and Luccioni, Sasha},
      year={2026},
      url={https://huggingface.co/spaces/hfmlsoc/different-flops},
}

@misc{grostern2026,
  author =  "Joey Grostern",
  year =  "2026",
  title =  "Data centres in Germany planning own fossil fuel plants, citing slow grid connections – report",
  date = "July",
  journal = "Clean Energy Wire",
  url =  "https://www.cleanenergywire.org/news/data-centres-germany-planning-own-fossil-fuel-plants-citing-slow-grid-connections-report",
}

@techreport{OpenAI-2026,
    author      =   "OpenAI",
    title       =   "The Hugging Face incident and the road ahead",
    institution =   "OpenAI",
    year        =   "2026", 
    note        =   "\url{https://openai.com/index/hugging-face-incident-and-the-road-ahead/}",
}

@techreport{Greenblatt-et-al-2026,
    author      =   "Ryan Greenblatt and Ajeya Cotra and Hjalmar Wijk",
    title       =   "Hugging Face incident investigation report",
    institution =   "METR",
    year        =   "2026",
    note        =   "\url{https://metr.org/blog/2026-08-26-openai-hugging-face-incident-investigation/}",
}

@misc{metr-2025,
    title = {Measuring the Impact of Early-2025 AI on Experienced Open-Source Developer Productivity},
    author = {Joel Becker and Nate Rush and Beth Barnes and David Rein},
    howpublished = {\url{https://metr.org/blog/2025-07-10-early-2025-ai-experienced-os-dev-study/}},
    year = {2025},
    month = {07},
}

@misc{ibm2026,
  author =       "Dave Bergmann",
  year =         "2026",
  title =        "What is tokenmaxxing?",
  date = "1 September",
  journal = "IBM Blog",
  url =  "https://www.ibm.com/think/topics/tokenmaxxing",
}

@misc{hausfather2026,
  author =       "Zeke Hausfather",
  year =         "2026",
  month =        "August",
  title =        "The real energy use of agentic AI",
  date = "August 05, ",
  url =  "https://www.theclimatebrink.com/p/the-real-energy-use-of-agentic-ai",
}

@misc{nyt2026,
  author =       "Mujib Mashal and Hari Kumar",
  year =         "2026",
  month =        "August",
  title =        "India Is Moving Fast to Catch Up in A.I. A Coastal City Fears the Fallout",
  date = "July 16th, ",
  journal = "New York Times",
  url =  "https://www.nytimes.com/2026/07/16/world/asia/india-ai-data-centers-google.html",
}

@misc{mittr2026,
  author =       "William Douglas Heaven",
  year =         "2026",
  month =        "March",
  title =        "I checked out one of the biggest anti-AI protests yet",
  date = "March 2, ",
  journal = "MIT Technology Review",
  url =  "https://www.technologyreview.com/2026/03/02/1133814/i-checked-out-londons-biggest-ever-anti-ai-protest/"
}

@misc{chen2026,
  author =       "Te-Ping Chen",
  year =         "2026",
  month =        May,
  title =        "Writers Are Going to Extremes to Prove They Didn’t Use AI",
  date = "May 6, ",
  journal = "The Wall Street Journal",
  url =  "https://www.wsj.com/tech/ai/writers-are-going-to-extremes-to-prove-they-didnt-use-ai-46e7c3f7"
}

@misc{graham2026,
  author =       "Michelai Graham",
  year =         "2026",
  month =        May,
  title =        "Your Next Favorite Bestseller May Be Written By AI. Why You Should Worry",
  date = "August 20, ",
  journal = "Ebony Magazine",
  url =  "https://www.ebony.com/next-favorite-bestseller-maybe-written-by-ai-why-should-worry/"
}

@misc{clarke2023,
  author =       "Neil Clarke",
  year =         "2023",
  month =        "February",
  title =        "A Concerning Trend",
  date = "February 15 ",
  journal = "Clarkesworld Magazine",
  url =  "https://neil-clarke.com/a-concerning-trend/"
}

@misc{dolezal2026,
      title={The Impact of AI-Generated Text on the Internet}, 
      author={Jonas Dolezal and Sawood Alam and Mark Graham and Maty Bohacek},
      year={2026},
      eprint={2604.26965},
      archivePrefix={arXiv},
      primaryClass={cs.CY},
      url={https://arxiv.org/abs/2604.26965}, 
}

@misc{epsteingross2023,
  author =       "Casey Epstein-Gross",
  year =         "2023",
  month =        "July",
  title =        "OpenAI Abruptly Shuts Down ChatGPT Plagiarism Detector—And Educators Are Worried",
  date = "July 26,",
  journal = "The Observer",
  url =  "https://observer.com/2023/07/openai-shut-ai-classifier/"
}

@article{liang2023gpt,
  title={GPT detectors are biased against non-native English writers},
  author={Liang, Weixin and Yuksekgonul, Mert and Mao, Yining and Wu, Eric and Zou, James},
  journal={Patterns},
  volume={4},
  number={7},
  year={2023},
  publisher={Elsevier}
}

@misc{brown2026,
  author =       "T.M. Brown",
  year =         "2026",
  month =       "October",
  title =        "Authors fight back against AI accusations and faulty detection tools",
  date = "Oct 21, ",
  journal = "CNN",
  url =  "https://www.wrtv.com/news/entertainment-news/authors-fight-back-against-ai-accusations-and-faulty-detection-tools/"
}

@misc{zhou2023,
  author =       "Viola Zhou",
  year =         "2023",
  month =        "April",
  title =        "AI is already taking video game illustrators’ jobs in China",
  date = "11 April",
  journal = "Rest of World",
  url =  "https://restofworld.org/2023/ai-china-video-game-layoffs-illustrators/"
}

@misc{reuters2026,
  author =       "Layli Foroudi and Forrest Crellin",
  year =         "2026",
  month =        "March",
  title =        "A backlash against data centres is spilling into French municipal election races",
  date = "March 13, ",
  journal="Reuters",
  url =  "https://www.reuters.com/sustainability/climate-energy/backlash-against-data-centres-is-spilling-into-french-municipal-election-races-2026-03-13/"
}

@misc{reuters2026_3,
  author =       "Reuters",
  year =         "2026",
  title =        "Companies cutting jobs as investments shift toward AI",
  date = "October 06,",
  journal="Reuters",
  url =  "https://www.reuters.com/business/world-at-work/companies-cutting-jobs-investments-shift-toward-ai-2026-10-06/"
}

@misc{atlantic2026,
  author =       "Ian Bogost",
  year =         "2026",
  month =        May,
  title =        "Why College Students Are Booing AI",
  date = "May 23, ",
  journal = "The Atlantic",
  url = "https://www.theatlantic.com/ideas/2026/05/ai-graduation-speeches-booing/687266/"
}

@inproceedings{Luccioni_2025, 
series={FAccT ’25},
   title={From Efficiency Gains to Rebound Effects: The Problem of Jevons’ Paradox in AI’s Polarized Environmental Debate},
   url={http://dx.doi.org/10.1145/3715275.3732007},
   DOI={10.1145/3715275.3732007},
   booktitle={Proceedings of the 2025 ACM Conference on Fairness, Accountability, and Transparency},
   publisher={ACM},
   author={Luccioni, Alexandra Sasha and Strubell, Emma and Crawford, Kate},
   year={2025},
   month="June", pages={76–88},
   collection={FAccT ’25} }

@misc{kim2026costdynamicreasoningdemystifying,
      title={The Cost of Dynamic Reasoning: Demystifying AI Agents and Test-Time Scaling from an AI Infrastructure Perspective}, 
      author={Jiin Kim and Byeongjun Shin and Jinha Chung and Minsoo Rhu},
      year={2026},
      eprint={2506.04301},
      archivePrefix={arXiv},
      primaryClass={cs.LG},
      url={https://arxiv.org/abs/2506.04301}, 
}

@misc{time2025,
  author =       "Billy Perrigo and Tharin Pillay",
  year =         "2025",
  month =        "October",
  title =        "‘Time Is Running Out’: New Open Letter Calls for Ban on Superintelligent AI Development",
  journal = "Time",
  date = "Oct 22, ",
  url =  "https://time.com/7327409/ai-agi-superintelligent-open-letter",
}

@misc{gazette2026,
  author =       "Montreal Gazette",
  year =         "2026",
  month =        "September",
  title =        "Photos: Protesters hit the streets during AI industry event in Montreal",
  journal = "Montreal Gazette",
  date = "September 16, ",
  url =  "https://montrealgazette.com/news/photos-anti-ai-protest-in-montreal/",
}

@misc{saig2026cleer,
    title        = {{CLEER}: Closed-model Latent Energy Estimation Range. Technical Report},
    author       = {Jegham, Nidhal and Koh, Chan Young and Gamazaychikov, Boris and Luccioni, Sasha},
    howpublished = {Sustainable AI Group},
    year         = {2026},
    month        = "sep",
    note         = {Version CLEER-Text-0926},
    url          = {https://reports.sustainableaigroup.com/CLEER-Tech-Report/}
  }

@misc{epoch2024computetrendpost2010,
title={The training compute of notable AI models has been doubling roughly every six months},
author={Robi Rahman and David Owen},
year={2024},
url={https://epoch.ai/data-insights/compute-trend-post-2010},
note={Accessed: 2026-10-16}}

@inproceedings{ai_environment_primer,
  author    = {Sasha Luccioni and
               Bruna Trevelin and
               Margaret Mitchell},
  title     = {The Environmental Impacts of AI -- Policy Primer},
  booktitle = {Hugging Face Blog},
  year      = {2024},
  url       = {https://doi.org/10.57967/hf/3004},
  doi       = {10.57967/hf/3004}
}

@article{kaplan2020scaling,
  title={Scaling laws for neural language models},
  author={Kaplan, Jared and McCandlish, Sam and Henighan, Tom and Brown, Tom B and Chess, Benjamin and Child, Rewon and Gray, Scott and Radford, Alec and Wu, Jeffrey and Amodei, Dario},
  journal={arXiv preprint arXiv:2001.08361},
  year={2020}
}

@article{Quintarelli-2019,
    author      =   "Stefano Quintarelli",
    title       =   "Let's forget the term {AI}. Let’s call them {Systematic Approaches to Learning Algorithms and Machine Inferences} ({SALAMI})",
    journal     =   "Quinta's Weblog",
    volume      =   "Nov.",
    number      =   "24",
    year        =   "2019",
    pages       =   "n.p.",
    note        =   "\url{https://blog.quintarelli.it/2019/11/lets-forget-the-term-ai-lets-call-them-systematic-approaches-to-learning-algorithms-and-machine-inferences-salami/}",
}

@article{Sandywell-2006,
    author      =   "Barry Sandywell",
    title       =   "Monsters in cyberspace: Cyberphobia and cultural panic in the information age",
    journal     =   "Information, Communication \& Society",
    volume      =   "9",
    number      =   "1",
    year        =   "2006",
    pages       =   "39--61",
}

@book{Wiener-1948,
    author      =   "Norbert Wiener",
    title       =   "Cybernetics: Or Control and Communication in the Animal and the Machine",
    publisher   =   "The Technology Press; John Wiley \& Sons",
    address     =   "New York",
    year        =   "1948",
}

@article{Global-National-2026,
    author      =   "{Global News}",
    title       =   "AI could `kill all humans' within next decade, researchers warn",
    journal     =   "Global National",
    volume      =   "Sep.",
    number      =   "09",
    year        =   "2026", 
    note        =   "\url{https://www.youtube.com/watch?v=4o2ribZ-AaU&t=51s}",
    pages       =   "n.p.",
}

@article{The-Guardian-2026-Control,
    author      =   "Garrison Lovely",
    title       =   "We have started losing control of AI. It's time to shut it down",
    journal     =   "The Guardian",
    volume      =   "Sep.",
    number      =   "10",
    year        =   "2026", 
    note        =   "\url{https://www.theguardian.com/commentisfree/2026/sep/10/ai-control-sci-fi}",
    pages       =   "n.p.",
}

@article{The-Guardian-2026-Bad,
    author      =   "Moustafa Bayoumi",
    title       =   "Sam Altman says AI will make `bad things' happen",
    journal     =   "The Guardian",
    volume      =   "Oct.",
    number      =   "08",
    year        =   "2026", 
    note        =   "\url{https://www.theguardian.com/commentisfree/2026/oct/08/ai-death-war-sam-altman}",
    pages       =   "n.p.",
}

@article{The-Guardian-2026-Letters,
    author      =   "Letters",
    title       =   "Could AI pose a serious threat to our existence?",
    journal     =   "The Guardian",
    volume      =   "Oct.",
    number      =   "08",
    year        =   "2026", 
    note        =   "\url{https://www.theguardian.com/technology/2026/sep/18/could-ai-pose-a-serious-threat-to-our-existence}",
    pages       =   "n.p.",
}

@article{Time-2026,
    author      =   "Geoffrey Irving",
    title       =   "We Won't Know the Answers to {AI}'s Most Important Questions Until It's Too Late",
    journal     =   "TIME",
    volume      =   "Oct.",
    number      =   "03",
    year        =   "2026", 
    note        =   "\url{https://time.com/article/2026/10/03/we-won-t-know-the-answers-to-ai-s-most-important-questions-until-its-too-late/}",
    pages       =   "n.p.",
}

@article{Reuters-2026,
    author      =   "Echo Wang and Aditya Soni",
    title       =   "Anthropic warns AI may pose 'existential risks to humanity' in IPO filing",
    journal     =   "Reuters",
    volume      =   "Sep.",
    number      =   "29",
    year        =   "2026",
    note        =   "\url{https://www.reuters.com/business/finance/anthropic-warns-ai-may-pose-existential-risks-humanity-ipo-filing-2026-09-29/}",
    pages       =   "n.p.",
}

@book{benjamin2024,
  title={Imagination: A manifesto},
  author={Benjamin, Ruha},
  year={2024},
  publisher={WW Norton \& Company}
}

\clearpage

\appendix

\section{`F*ck AI' Gallery}
\label{sec:A}

\phantom{a}

\begin{figure}[htb!]
    \centering
        \includegraphics[width=0.45\linewidth]%
        {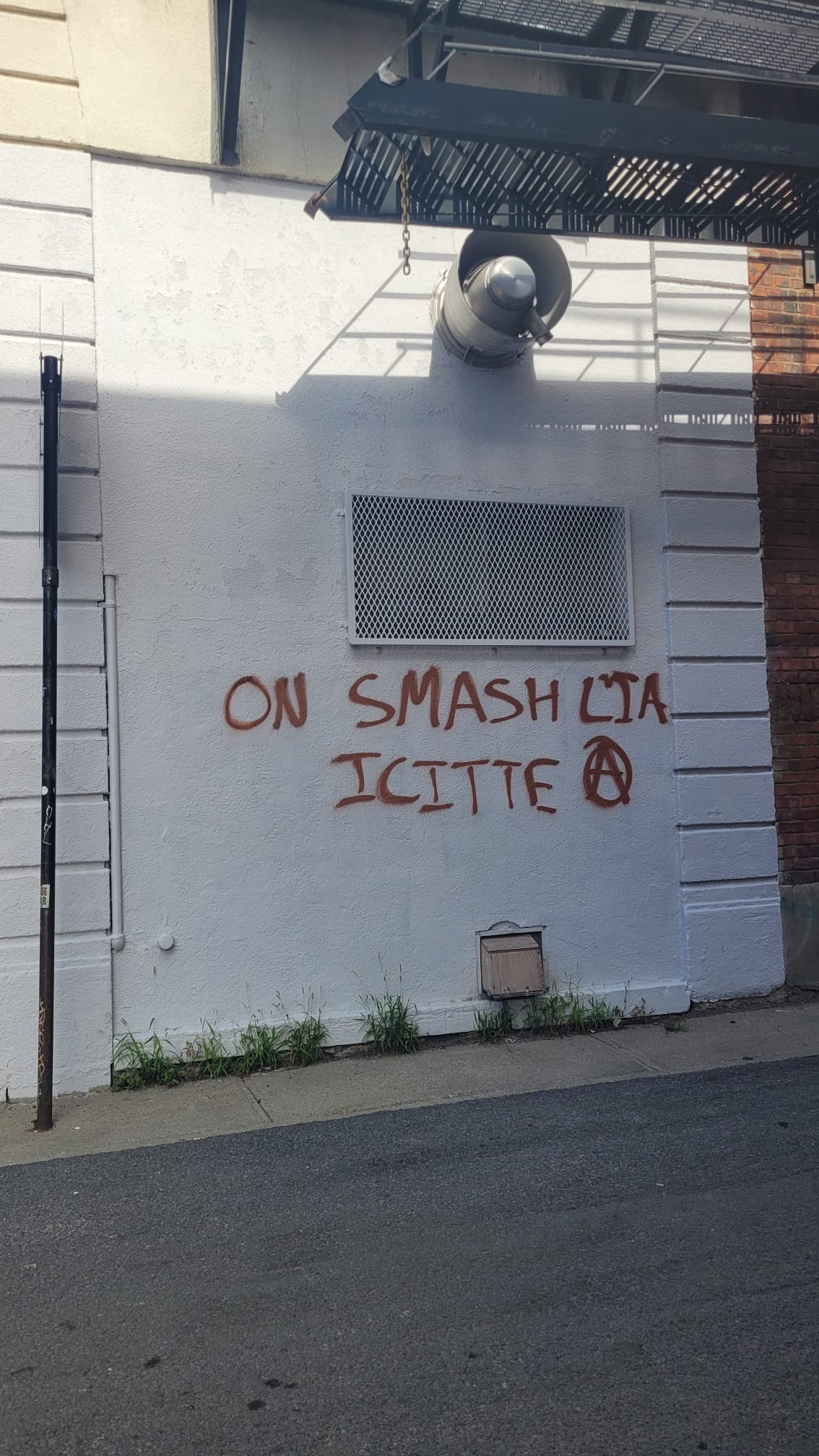}~
        \includegraphics[width=0.45\linewidth]%
        {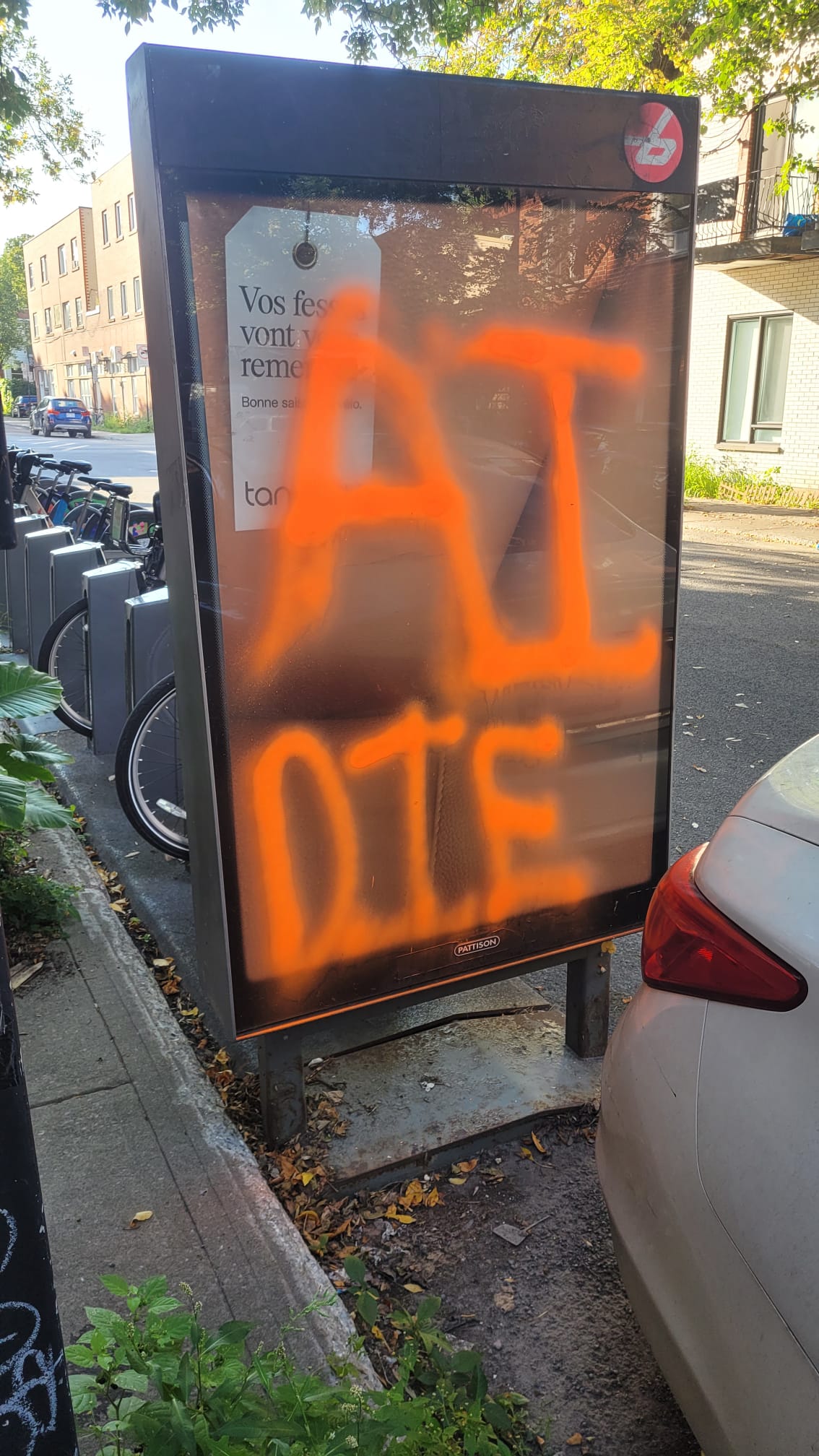}
        \caption{Graffiti reading ``ON SMASH L'IA ICITTE'' [``we smash AI here''] along with the circle-A (\CircledA), written on the wall of a Montr{\'e}al building (left), and ``AI DIE'' (right), written over an advertising column at a bike-share station in Montr{\'e}al; photos taken by Sasha Luccioni.}
    \Description{}
    \label{fig:Graffiti-M1}
\end{figure}

\newpage
\begin{figure}[htb!]
    \centering
        \includegraphics[width=0.4\linewidth]%
        {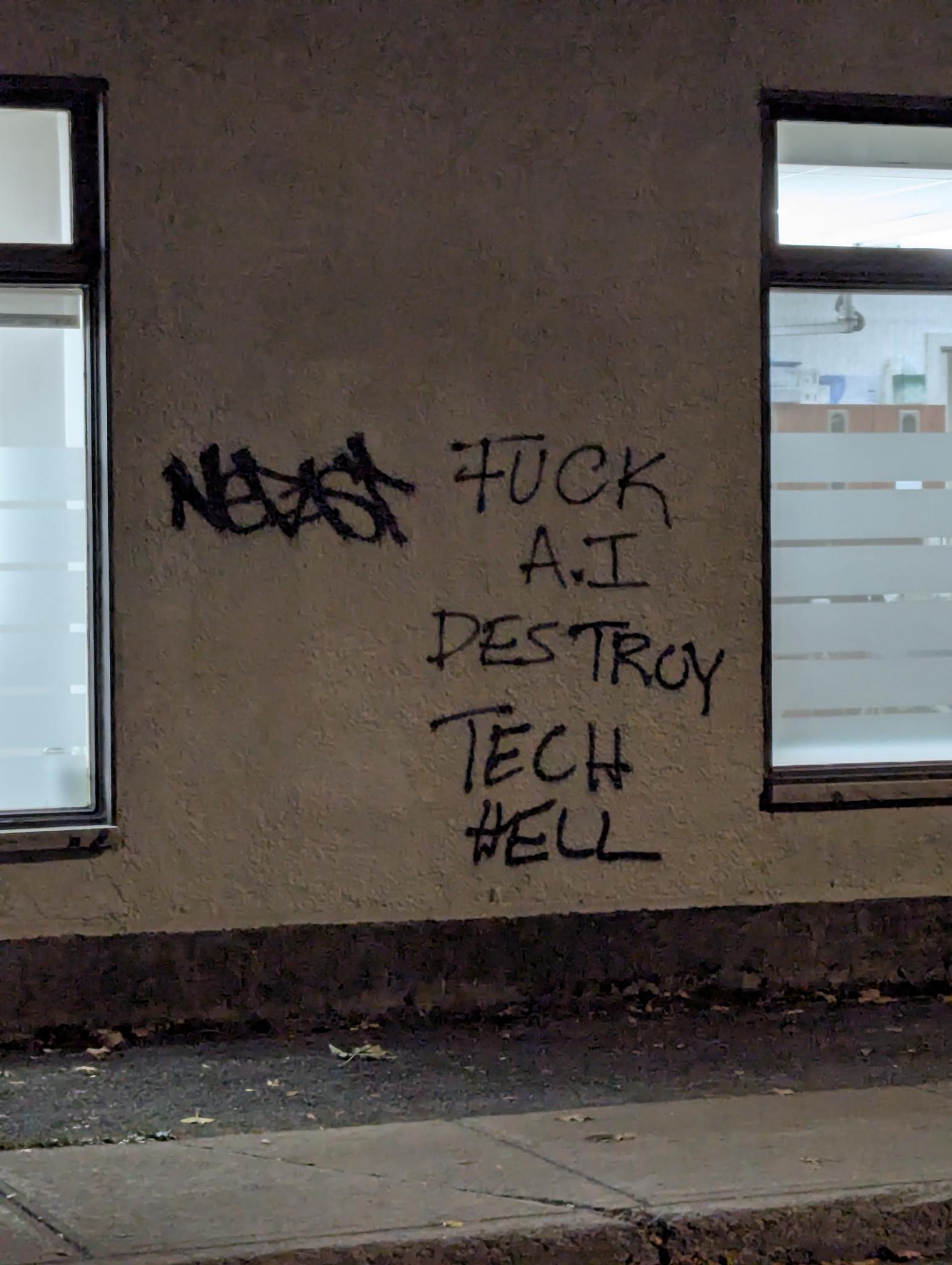}~
        \includegraphics[width=0.4\linewidth]%
        {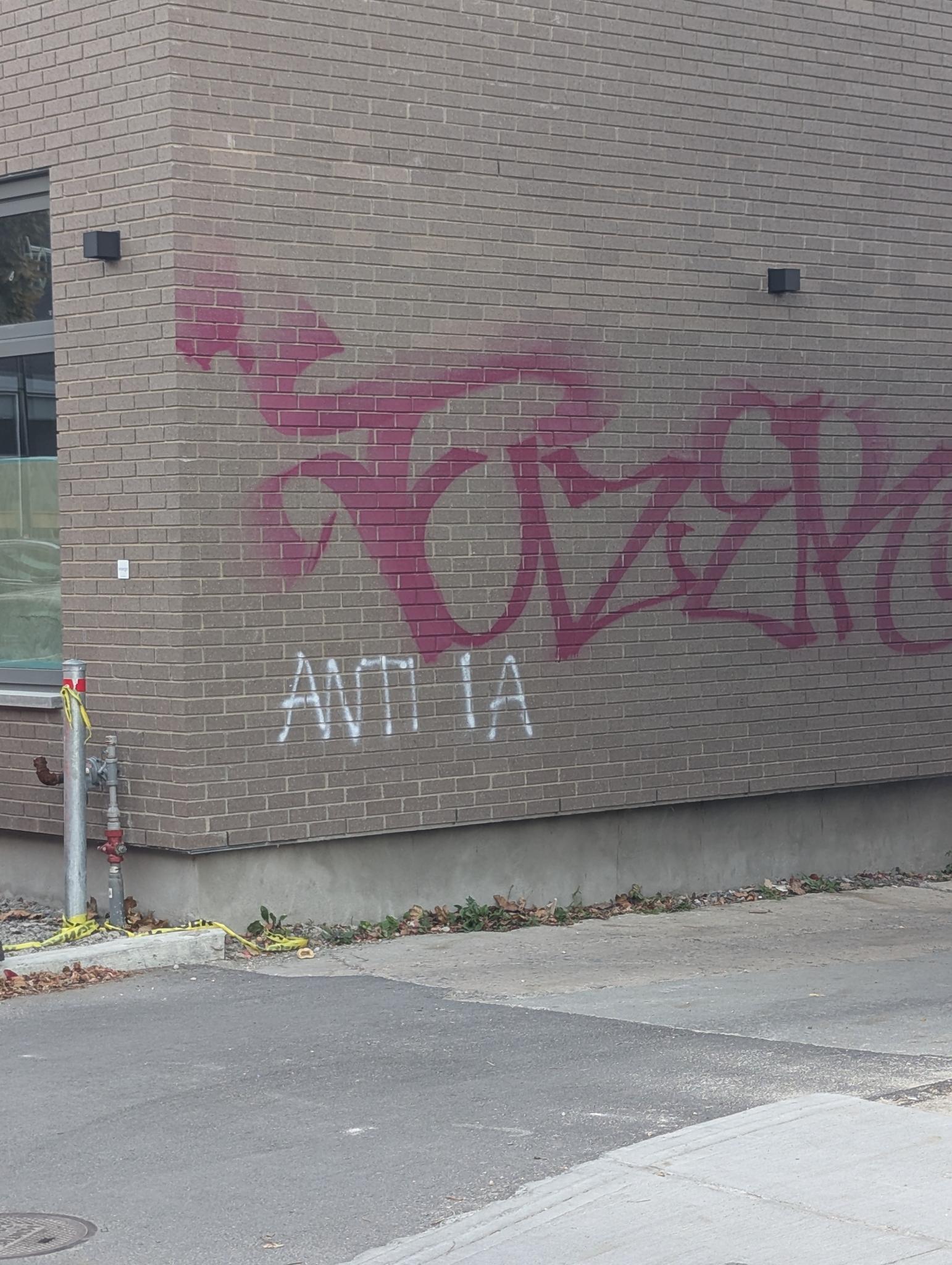} 
        \includegraphics[width=0.4\linewidth]%
        {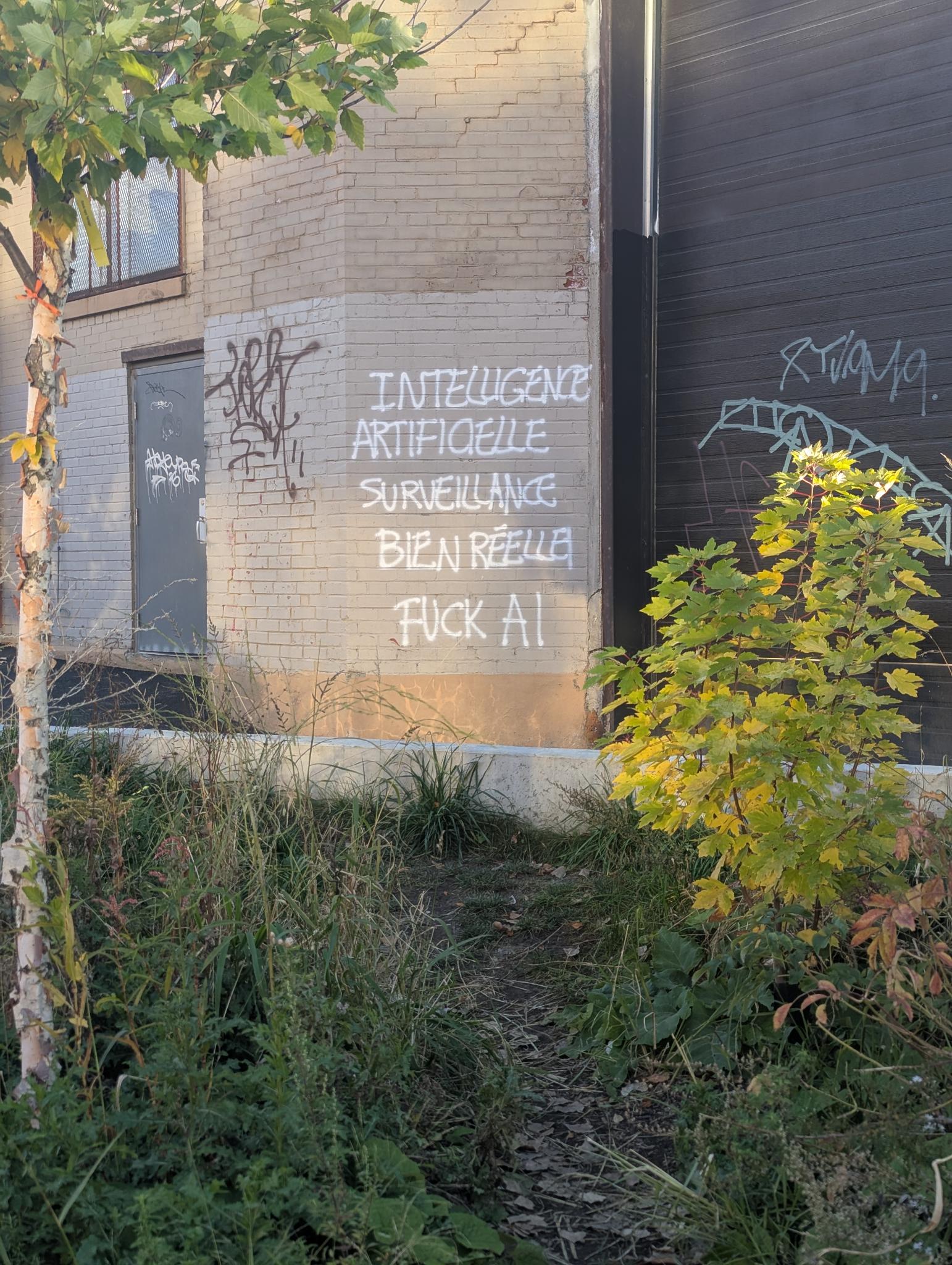}~
        \includegraphics[width=0.4\linewidth]%
        {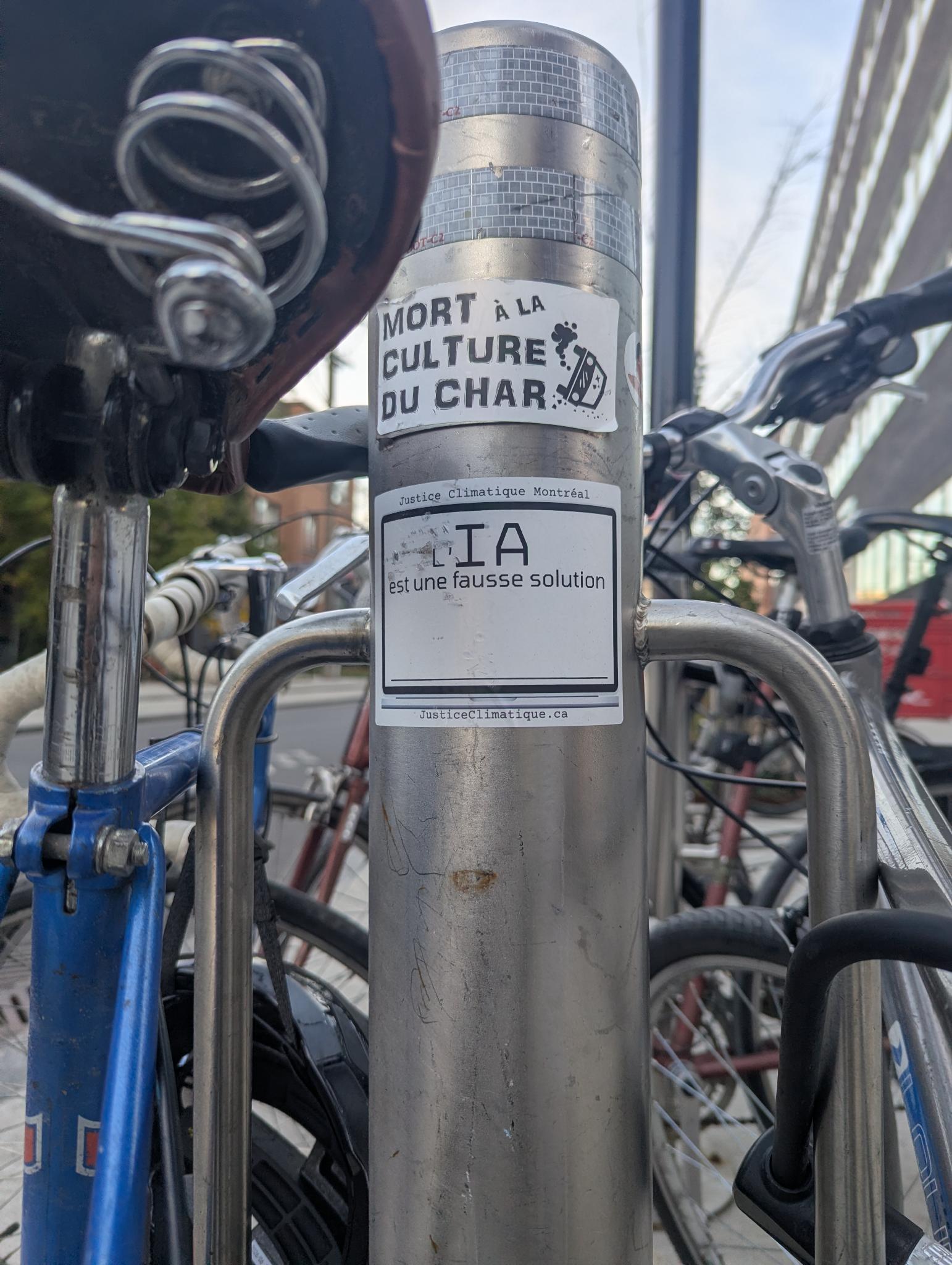}
        \caption{Graffiti reading ``FUCK AI DESTROY TECH HELL'' (top left), ``ANTI IA'' [``Anti AI''] (top right), ``INTELLIGENCE ARTIFICIELLE SURVEILLANCE BIEN R{\'E}ELLE FUCK AI'' [``artificial intelligence, very real surveillance; Fuck AI''] (bottom left), written on the walls of various Montr{\'e}al buildings, and a sticker reading ``IA est une fausse solution'' [``AI is a false solution''], on a bike rack; photos taken by Michael Noukhovitch.
        }
    \Description{}
    \label{fig:Graffiti-M1}
\end{figure}

\newpage

\begin{figure}[htb!]
    \centering
        \includegraphics[width=0.608\linewidth]%
        {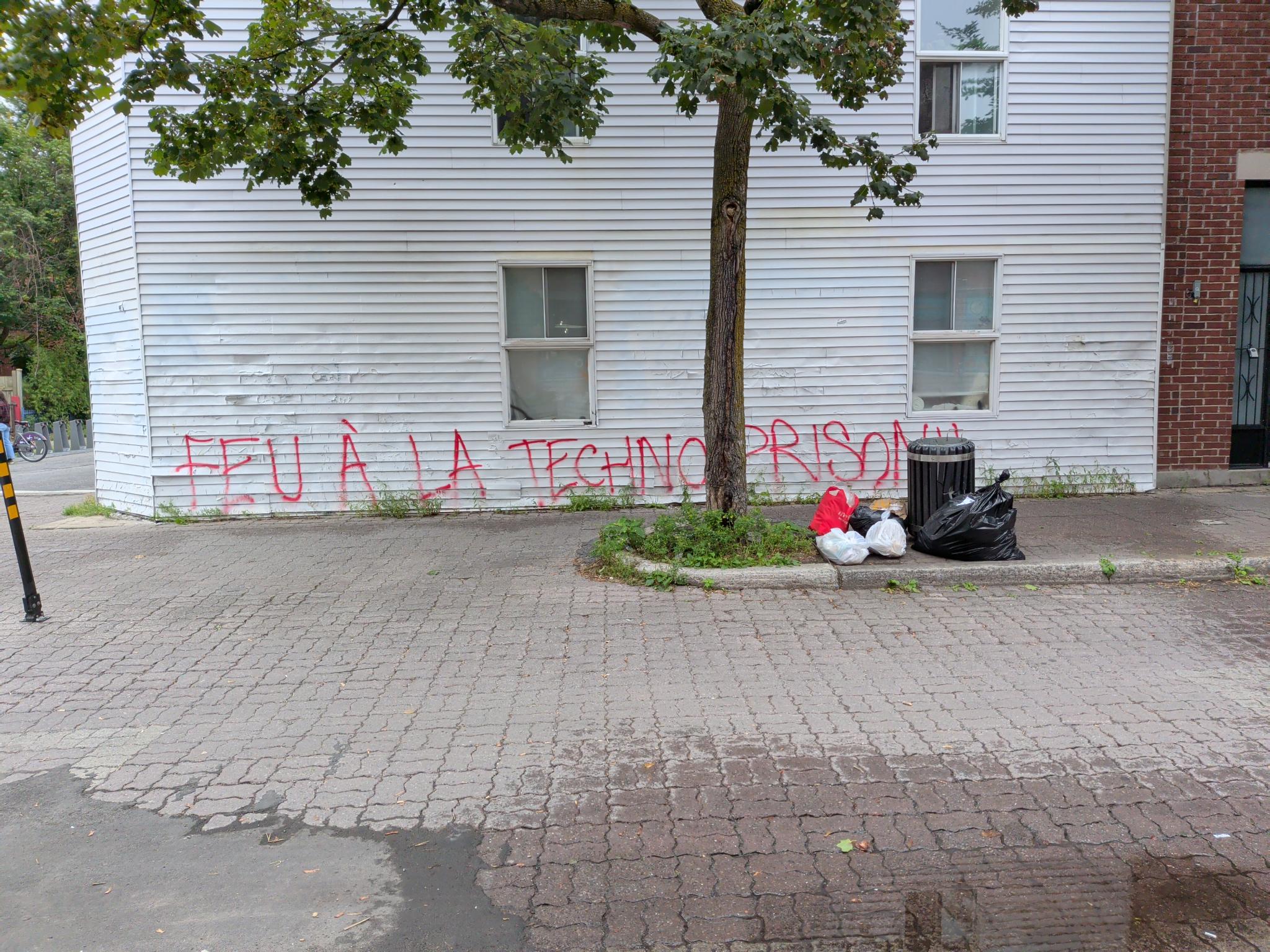}~
        \includegraphics[width=0.342\linewidth]%
        {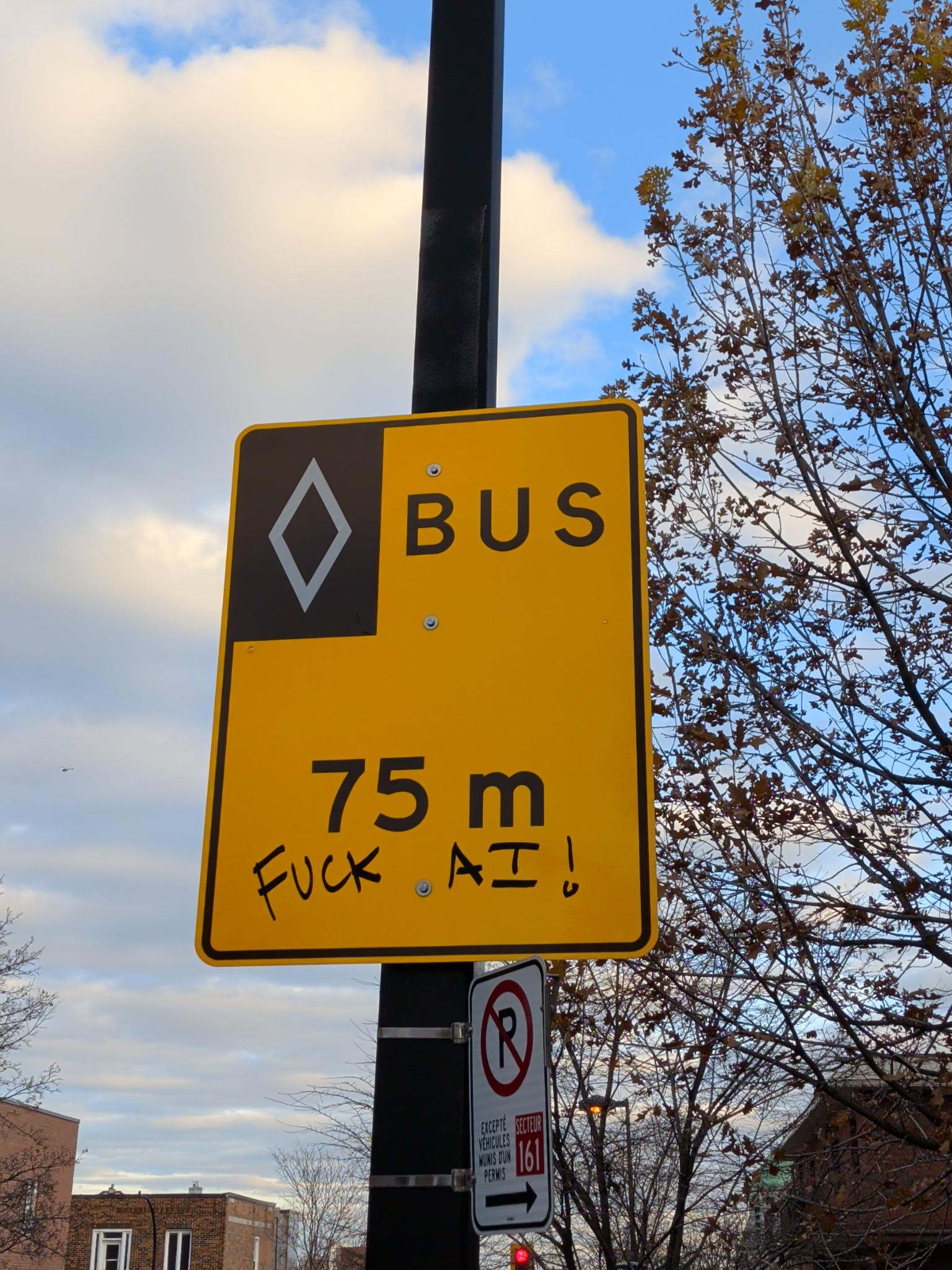}
        \caption{Graffiti reading ``FEU {\`A} LA TECHNO PRISON'' [``Fire at the techno prison''], written on the wall of a Montr{\'e}al building (left), and ``FUCK AI'' (right), written over a traffic sign indicating a bus lane in 75 m (Montr{\'e}al); photos taken by Travis LaCroix (left) and Lynn LaCroix (Right).}
    \Description{}
    \label{fig:Graffiti-M1}
\end{figure}

\end{document}